\documentclass[a4paper,11pt]{article}
\pdfoutput=1 % if your are submitting a pdflatex (i.e. if you have
\usepackage{jheppub} % for details on the use of the package, please
\usepackage[utf8]{inputenc}                      
\usepackage{subfig}
\usepackage[T1]{fontenc} % if needed
\usepackage{graphicx}
\usepackage{epsfig}
\usepackage{axodraw}
\usepackage{amsmath}
\usepackage{amssymb}
\usepackage{float}
\usepackage{placeins}
\usepackage[utf8]{inputenc}
\usepackage{changepage}
\usepackage{bm}
\usepackage{multicol}
\usepackage{comment}
\usepackage[font=small,labelfont=bf]{caption}

\newcommand{\imgBotRe}[1]{\includegraphics[scale=0.25]{figures/plot/plot4_Bottom_Re_#1.png}}
\newcommand{\imgBotIm}[1]{\includegraphics[scale=0.25]{figures/plot/plot4_Bottom_Im_#1.png}}
\newcommand{\imgTopRe}[1]{\includegraphics[scale=0.25]{figures/plot/plot4_Top_Re_#1.png}}
\newcommand{\imgTopIm}[1]{\includegraphics[scale=0.25]{figures/plot/plot4_Top_Im_#1.png}}
\newcommand{\icol}[1]{% inline column vector
  \left(\begin{smallmatrix}#1\end{smallmatrix}\right)%
}

\allowdisplaybreaks[4]

\title{The complete set of two-loop master integrals for Higgs $+$ jet production in QCD}

\preprint{MPP-2019-228\\
	\phantom{~} \hfill  TCDMATH 19-29}

\author[a,b]{H.~Frellesvig,}\author[c,d]{M.~Hidding,}
\author[e]{L.~Maestri,}\author[f]{F.~Moriello,}\author[g,h,i]{G.~Salvatori}

\affiliation[a]{Dipartimento di Fisica e Astronomia, Universit{\`a} di Padova, Via Marzolo 8, 35131 Padova, Italy}

\affiliation[b]{INFN, Sezione di Padova, Via Marzolo 8, 35131 Padova, Italy}

\affiliation[c]{Hamilton Mathematics Institute, Trinity College, Dublin 2, Ireland}

\affiliation[d]{School of Mathematics, Trinity College, Dublin 2, Ireland}

\affiliation[e]{Max-Planck-Institut f\"ur Physik, Werner-Heisenberg-Institut, D-80805 M\"unchen, Germany}

\affiliation[f]{ETH Zurich, Institut f{\"u}r theoretische Physik, Wolfgang-Paulistr. 27, 8093, Zurich, Switzerland}

\affiliation[g]{Department of Physics, Brown University, Providence, RI 02912, USA}

\affiliation[h]{INFN, Sezione di Milano, Via Celoria 16, 20133 Milano, Italy}

\affiliation[i]{Dipartimento di Fisica, Universit{\`a} degli Studi di Milano, Via Celoria 16, 20133 Milano, Italy}

\emailAdd{hjalte.frellesvig@pd.infn.it}
\emailAdd{hiddingm@tcd.ie}
\emailAdd{maestri@mpp.mpg.de}
\emailAdd{fmoriell@phys.ethz.ch}
\emailAdd{giulio$\_$salvatori@brown.edu}

\abstract{In this paper we complete the computation of the two-loop master integrals relevant for Higgs plus one jet production initiated in~\cite{Bonciani:2016qxi,Bonciani:2019jyb,Francesco:2019yqt}. Specifically, we compute the remaining family of non-planar master integrals. The computation is performed by defining differential equations along contours in the kinematic space, and by solving them in terms of one-dimensional generalized power series. This method allows for the efficient evaluation of the integrals in all kinematic regions, with high numerical precision. We show the generality of our approach by considering both the top- and the bottom-quark contributions. This work along with~\cite{Bonciani:2016qxi,Bonciani:2019jyb,Francesco:2019yqt} provides the full set of master integrals relevant for the NLO corrections to Higgs plus one jet production, and for the real-virtual contributions to the NNLO corrections to inclusive Higgs production in QCD in the full theory.}

\begin{document}

\maketitle

\section{Introduction}
The main production mode of the Higgs boson at the Large Hadron Collider (LHC) is via gluon fusion. In perturbative Quantum Chromo Dynamics (QCD) the production is mediated by a quark loop  that couples to the final-state Higgs. The quark-Higgs coupling is proportional to the quark mass, hence the largest contribution is given by corrections involving a top-quark. Being mediated by a quark loop, the leading-order (LO) corrections require the computation of one-loop amplitudes while the next-to-leading-order (NLO) corrections require the computation of two-loop amplitudes and so on. The inclusive LO corrections to Higgs production have been computed in the full theory at LO in \cite{Georgi:1977gs} and at NLO in \cite{Graudenz:1992pv,Spira:1995rr}. On the other hand, the computation of the higher order corrections is much more challenging, and complete results in the full theory are not yet available. The computation can be considerably simplified in the limit where the top quark is assumed to be infinitely heavy, while the other quarks are assumed to be massless. This limit is known as the Higgs Effective Field Theory (HEFT). The next-to-next-to-leading-order (NNLO) QCD corrections have been computed in the HEFT in \cite{Harlander:2002wh,Anastasiou:2002yz,Ravindran:2003um} while, more recently, the corresponding next-to-next-to-next-to-leading-order (N$^3$LO) corrections have been computed in~\cite{Anastasiou:2015ema,Mistlberger:2018etf}. 

In addition to inclusive cross-sections, differential cross sections play an important role in the study of the properties of the Higgs boson. In particular, the Higgs may couple to particles not predicted by the Standard Model, and many such effects will be best studied by observing the transverse momentum ($p_T$) distribution of the Higgs~\cite{Harlander:2013oja,Banfi:2013yoa,Azatov:2013xha,Grojean:2013nya,Schlaffer:2014osa,Buschmann:2014twa,Dawson:2014ora,Buschmann:2014sia,Ghosh:2014wxa,Dawson:2015gka,Langenegger:2015lra,Azatov:2016xik,Grazzini:2016paz,Grazzini:2018eyk,Gorbahn:2019lwq}, particularly at high $p_T$. In the full theory the Higgs plus jet production cross section and the $p_T$ distribution are only known at LO. At NLO, the top-quark contributions have been computed in ~\cite{Jones:2018hbb}, while the top-bottom interference was computed in~\cite{Lindert:2017pky} by combining the HEFT with an asymptotic expansion around small bottom-mass. At higher perturbative order no result is available in the full theory, and only partial results are known in the HEFT. More specifically, the NNLO corrections to the Higgs plus one jet production and the Higgs $p_T$ distribution are known in the HEFT. However, while the HEFT approximation works well for inclusive observables, it diverges very rapidly for high-energy differential observables, such as the high $p_T$ distribution of the Higgs (see e.g.~\cite{Neumann:2016dny} and references therein).

To this date no complete result for the Higgs plus jet amplitudes at NLO is available in the full theory. The first step in this direction has been taken in \cite{Bonciani:2016qxi} and, more recently, in \cite{Francesco:2019yqt,Bonciani:2019jyb}, where the planar master integrals and one of the two non-planar families of master integrals at two loops have been computed in terms of one-dimensional generalized power series. This technique is not constrained in any way to a particular kinematic region or a specific configuration of the relevant masses, and allows for the efficient computation of the master integrals while keeping the full dependence on all the mass scales. In this paper, we apply this technique to compute the remaining family of non-planar master integrals. Besides the NLO QCD corrections to Higgs plus jet production, these master integrals are an ingredient of the NLO corrections to Higgs decay to three partons, and are also a building block of the NNLO inclusive corrections to Higgs production in the full theory, where the Higgs plus jet amplitudes appear as the single real radiation contribution.

The paper is organised as follows.  In section \ref{sec:integral family} we define the non-planar integral family computed in this paper. In section \ref{sec:DE} we review the differential equations method for dimensionally regulated Feynman integrals, and we discuss the structure of the differential equations of our integral family.  In section \ref{sec:series_exp} we describe our solution strategy, i.e. we solve differential equations along contours in the space of kinematic invariants in terms of generalized power series. In section \ref{sec:results} we show how the expansion strategy is used to evaluate the master integrals in a very large sample of points in the physical region, for both the top- and bottom-quark contributions. In section \ref{sec:concl} we draw our conclusions and we discuss directions for future work.

%\section{The non-planar integral families}
\section{The integral family}
\label{sec:integral family}

\begin{figure}[!ht]
\centering
\includegraphics[width=0.5\textwidth]{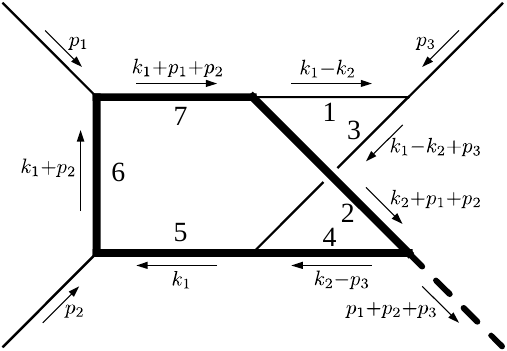}
\caption{The integral family with momenta and propagator labels.}
\label{fig:famdef}
\end{figure}

As discussed in ref.~\cite{Bonciani:2019jyb}, six seven-propagator integral families contribute to the two-loop QCD contribution to $H+jet$  production. Of these families, four are planar and have been computed in ref.~\cite{Bonciani:2016qxi}, and of the non-planar, one was computed in ref.~\cite{Bonciani:2019jyb} and the remaining one, denoted family $G$, will be the topic of the present paper.

That integral family is defined by
\begin{align}
\text{I}_{a_1 a_2 a_3 a_4 a_5 a_6 a_7 a_8 a_9} &= e^{2 \gamma_{E} \epsilon} \int \!\! \int \frac{d^D k_1 d^D k_2}{(i \pi^{d/2})^2} \frac{P_8^{-a_8} \, P_9^{-a_9}}{P_1^{a_1} P_2^{a_2} P_3^{a_3} P_4^{a_4} P_5^{a_5} P_6^{a_6} P_7^{a_7}}
\label{Is}
\end{align}
where $\gamma_{E}=-\Gamma^{\prime}(1)$ is the Euler-Mascheroni constant, and where
\begin{align}
P_1 &= -(k_1-k_2)^2, & P_4 &= m^2-(k_2{-}p_3)^2, & P_7 &= m^2 - (k_1{+}p_1{+}p_2)^2, \nonumber \\
P_2 &= m^2 -(k_2{+}p_1{+}p_2)^2, & P_5 &= m^2 - k_1^2, & P_8 &= m^2 - k_2^2, \\
P_3 &= -(k_1{-}k_2{+}p_3)^2, & P_6 &= m^2 - (k_1{+}p_2)^2, & P_9 &= - (k_1{-}k_2{-}p_1)^2. \nonumber
\end{align}
Only $P_1$-$P_7$ can appear as genuine propagators, so we have $a_8$ and $a_9$ restricted to the non-positive integers. The kinematics is $p_1^2 = p_2^2 = p_3^2 = 0$ and additionally
\begin{align}
s&\equiv (p_1{+}p_2)^2, & t &\equiv (p_1{+}p_3)^2, & u &\equiv (p_2{+}p_3)^2, & p_4^2 &= (p_1{+}p_2{+}p_3)^2 = s{+}t{+}u,
\end{align}
where $m^2$ denotes the squared mass of the quark that couples to the Higgs, and $p_4^2$ the squared mass of the Higgs.

By using integration-by-parts (IBP)~\cite{Tkachov:1981wb,Chetyrkin:1981qh,Laporta:1996mq,Laporta:2001dd} reduction methods \cite{Maierhoefer:2017hyi,Smirnov:2019qkx}, we identify a set of 84 master integrals for this family, whose diagrams are shown in Fig. \ref{fig:masters}. With those master integrals we defined a basis of Feynman integrals which is presented in Appendix \ref{app:CanonicalBasis}.

\begin{figure}[!h]
\centering
\includegraphics[width=0.82\textwidth]{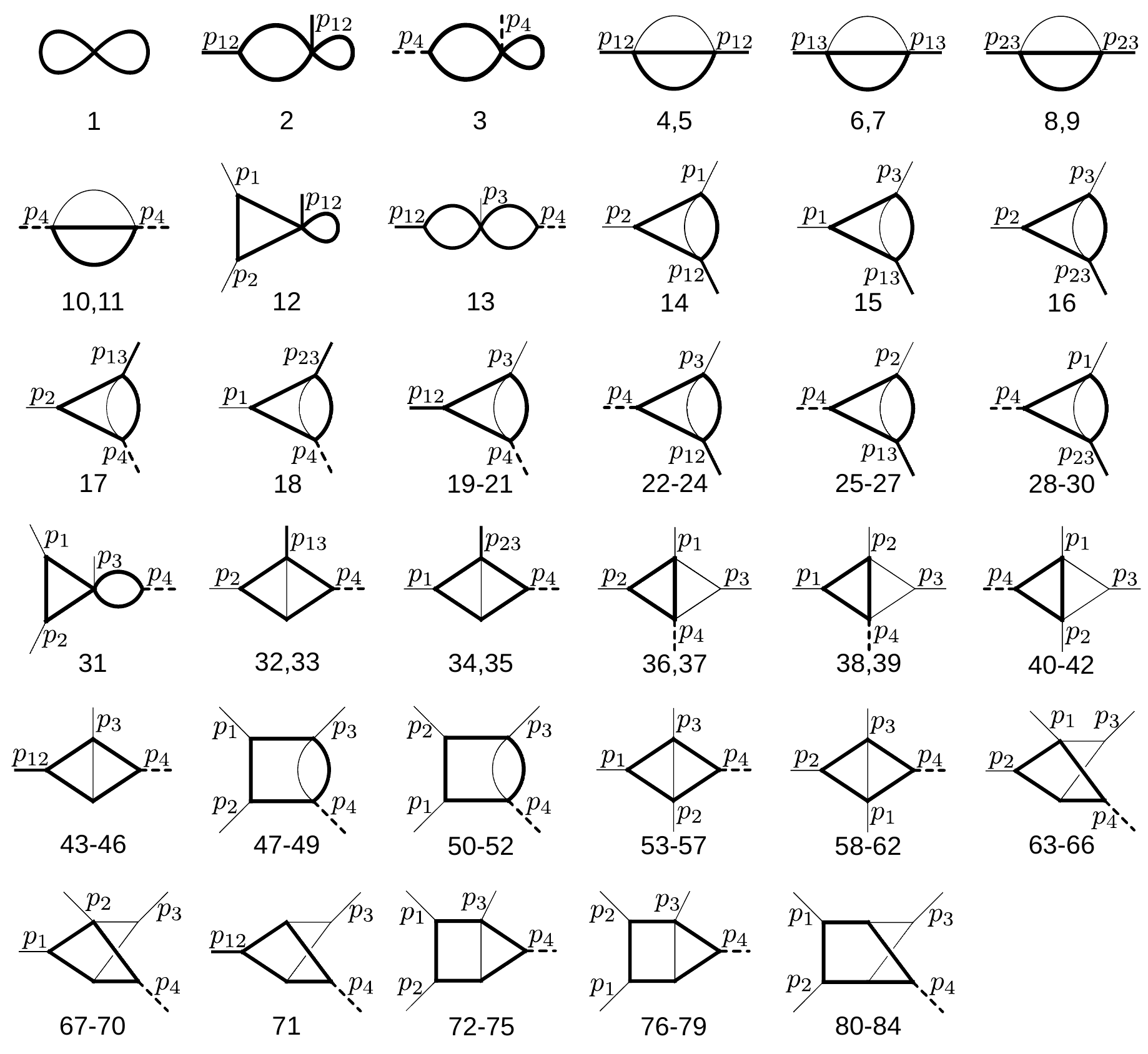}
\caption{The 84 master integrals. Shown on the figure is the sector, i.e. the set of propagators, to which the master integrals belong. Higher powers of propagators, numerators, or prefactors are not shown. External momenta are labelled using $p_{ij} = p_i{+}p_j$ and $p_4 = p_1{+}p_2{+}p_3$. Masses (internal as well as external) are indicated with a thicker line.}
\label{fig:masters}
\end{figure}

\section{Differential equations for the integral family}
\label{sec:DE}

Given a basis of $N$ master integrals $\vec{I}(\epsilon,\vec{s})$, where $\epsilon$ is the dimensional regulator defined by $D=4-2 \epsilon$ and $\vec{s}=\{s_1,\ldots,s_n\}$ is a set of $n$ Lorentz invariants, it is possible to define a closed system of linear, first order differential equations \cite{Kotikov:1990kg,Kotikov:1991pm,Bern:1993kr,Remiddi:1997ny,Gehrmann:1999as}  for $\vec{I}(\epsilon,s_i)$ that in full generality reads,
\begin{equation}
    \partial_{s_i}\vec{I}(\epsilon,\vec{s})=\mathbf{M}_{s_i}(\epsilon,\vec{s})\vec{I}(\epsilon,\vec{s}),
\end{equation}
where $\partial_{s_i}\equiv \frac{\partial}{\partial s_i}$ and  $\mathbf{M}_{s_i}$ is a set of $N\times N$ matrices.

The choice of the basis integrals is not unique, and by performing a basis change $\vec{B}=\mathbf{T} \vec{I}$ the differential equations transform according to,
\begin{equation}
    \label{eq:canformpartials1}
    \partial_{s_i}\vec{B} (\epsilon,\vec{s}) = \left(\mathbf{T}\mathbf{M}_{s_i}\mathbf{T}^{-1}(\epsilon,\vec{s}) - \mathbf{T}\partial_{s_i} \mathbf{T}^{-1}(\epsilon,\vec{s})\right) \vec{B}(\epsilon,\vec{s})\,. 
\end{equation}
In Ref.~\cite{Henn:2013pwa} it was conjectured that with a proper basis choice it is possible to cast the differential equations for Feynman integrals in the following simplified form,
\begin{equation}
    \label{eq:canonical}
    \partial_{s_i}\vec{B} (\epsilon,\vec{s}) = \epsilon \mathbf{A}_{s_i}(\vec{s}) \vec{B}(\epsilon,\vec{s})\,, 
\end{equation}
where the dependence on $\epsilon$ is factorised out, and the matrices $\mathbf{A}_{s_i}(\vec{s})$ depend only on the invariants $\vec{s}$. Such a system of differential equations is said to be in canonical form, and the basis $\vec{B}$ is referred to as the canonical basis. A canonical system of differential equations is equivalent to the following equation in differential form,
\begin{equation}
\label{eq:d_canonical}
    d\vec{B}(\epsilon,\vec{s})=\epsilon \,d\tilde{\mathbf{A}}(\vec{s})\vec{B}(\epsilon,\vec{s})\,,
\end{equation}
where, by construction, the matrix $\tilde{\mathbf{A}}$ satisfies
\begin{equation}
    \partial_{s_i}\tilde{\mathbf{A}}(\epsilon,\vec{s})=\mathbf{A}_{s_i}(\epsilon,\vec{s})\,.
\end{equation}

The differential equation~(\ref{eq:d_canonical}) can be formally solved in terms of a path-ordered exponential
\begin{equation}
    \vec{B}(\epsilon,\vec{s})=\mathcal{P}\exp\left(\epsilon \int_{\gamma} d\tilde{\mathbf{A}}(\vec{s})\right)\vec{B}(\epsilon,\vec{s}_0)\,,
\end{equation}
where $\gamma$ is an integration path connecting a boundary point $\vec{s}_0$ to $\vec{s}$, and $\mathcal{P}$ is the path-ordering operator. In dimensional regularization we are generally interested in a solution around $\epsilon=0$. By performing the expansion for small $\epsilon$, the path-ordered exponential translates to iterated integrals over the entries of $\tilde{\mathbf{A}}(\vec{s})$. Specifically, the solution to all orders of $\epsilon$ is
\begin{equation}
\label{eq:sol_iterated_integrals}
    \vec{B}(\epsilon,\vec{s})=\vec{B}(\epsilon,\vec{s}_0)+\sum_{k\geq 1}\epsilon^k \sum_{j=1}^k \int_{0\leq t_j\leq \ldots \leq t_1 \leq 1}  \gamma^*(d\tilde{\mathbf{A}}(t_1))   \ldots \gamma^*(d\tilde{\mathbf{A}}(t_j)) \,\vec{B}^{(k-j)}(\vec{s}_0)\,,
\end{equation}
where $\gamma: [0,1]\rightarrow \mathbb{R}^n $ and $\gamma^*(d\tilde{\mathbf{A}}(t_i))=\frac{\partial \tilde{\mathbf{A}}(t_i)}{\partial t_i} d t_i$, while $\vec{B}^{(i)}(\vec{s})$ denotes the $i$-th coefficient of the $\epsilon$-expansion.

So far we made no assumptions on the class of functions arising from the iterated integrals of Eq.~(\ref{eq:sol_iterated_integrals}). A large class of master integrals admits a canonical basis such that the matrix $\tilde{\mathbf{A}}(\vec{s})$ is a $\mathbb{Q}$-linear combination of logarithms of rational or algebraic arguments. This form also implies that the transformation $\mathbf{T}(\epsilon,\vec{s})$ to the canonical basis is rational or algebraic. The logarithms can be chosen in such a way that there are no $\mathbb{Q}$-linear relations between them. In the literature, the arguments of the independent logarithms are typically referred to as letters, while the set of letters is referred to as the alphabet.

If the alphabet contains only rational functions, the solutions can be directly expressed in terms of multiple polylogarithms~\cite{Goncharov:1998kja}, which are defined recursively as,
\begin{equation}
    G(a_1,a_2,\ldots,a_n,x)=\int_0^x\frac{dt}{t-a_1}G(a_2,\ldots,a_n,t),
\end{equation}
with $G(,x)\equiv 1$ and $G(\vec{0}_n,x)\equiv\frac{\log(x)^n}{n!}$.
On the other hand, the general case where the alphabet contains algebraic functions is much less understood. In some cases it is possible to rationalize the algebraic functions by a suitable reparametrization of the invariants, reducing in this way the problem to a rational one. If a rational parametrization is not available, it is possible, in some case, to define an ansatz for the solution in terms of polylogarithms of suitably chosen (algebraic) arguments. The unknown parameters of the ansatz are then fixed by solving the differential equations and by imposing boundary conditions (see e.g. \cite{Brown:2009qja, Goncharov.A.B.:2009tja, Goncharov:2010jf, Duhr:2011zq,Bonciani:2016qxi,Heller:2019gkq,Bonciani:2019jyb}). Nonetheless, in the general algebraic case, it is not known whether the differential equations always admit a solution in terms of multiple polylogarithms.

More recently, a lot of progress has been made in the study of Feynman integrals which evaluate to elliptic generalizations of multiple polylogarithms (eMPLs) \cite{BrownLevin,Broedel:2014vla,Broedel:2017kkb,Laporta:2004rb,Kniehl:2005bc,Adams:2013kgc,Bloch:2013tra,Bloch:2014qca,Adams:2014vja,Adams:2015gva,Adams:2015ydq,Remiddi:2016gno,Primo:2016ebd,Bonciani:2016qxi,Adams:2016xah,Passarino:2016zcd,Harley:2017qut,vonManteuffel:2017hms,Ablinger:2017bjx,Chen:2017pyi,Hidding:2017jkk,Bogner:2017vim,Bourjaily:2017bsb,Broedel:2017siw,Laporta:2017okg,Broedel:2018iwv,Mistlberger:2018etf,Lee:2018jsw,Broedel:2018qkq,Adams:2018bsn,Adams:2018kez,Broedel:2019hyg,Bogner:2019lfa,Kniehl:2019vwr,Broedel:2019kmn}. However, while in some cases it is possible to define a basis that casts the differential equations in canonical form (see e.g. \cite{Adams:2018yfj,Adams:2018bsn,Adams:2018kez,Broedel:2019kmn}), little is known about their general analytic properties and how to systematically solve them in terms of elliptic multiple polylogarithms.

As we will see in the next sections, the differential equations for the integral family considered in this paper depend on complicated algebraic functions. Moreover some equations are coupled, and their solution involves functions of elliptic type. In this case finding a closed form solution for the integrals seems to be out of reach with current technology.  Nonetheless, having phenomenological applications in mind, we follow a different approach, based on the series solution of the differential equations along contours in the space of the kinematic variables \cite{Francesco:2019yqt}. 

\subsection{Canonical integrals}
 We denote the set of Lorentz invariants as $\vec{s}=\{s_1,s_2,s_3,s_4\}=\{s,t,p_4^2,m^2\}$. The first 71 master integrals of the basis chosen for family G are such that the system of differential equations are in canonical form. Namely, they satisfy Eq. (\ref{eq:d_canonical}).
% canonical basis, and they satisfy differential equations of the form,
%\begin{equation}
%    \label{eq:canonicalG}
%    \partial_{s_i}\vec{B}_{1-71} (\epsilon,\vec{s}) = \epsilon \mathbf{A}_{s_i}(\vec{s}) \vec{B}_{1-71}(\epsilon,\vec{s})\,. 
%\end{equation}
%It is possible to cast these differential equations in the following d-log form,
%\begin{equation}
%\label{eq:d_canonicalG}
%    d\vec{B}_{1-71} (\epsilon,\vec{s})=\epsilon \,d\tilde{\mathbf{A}}(\vec{s})\vec{B}_{1-71} (\epsilon,\vec{s}),
%\end{equation}
The matrix $\tilde{\mathbf{A}}$ can be constructed by using the following iterative definitions,
\begin{equation}
\label{eq:Atildefrompdes}
\tilde{\mathbf{A}}_1 \equiv \int \mathbf{A}_{s_1} ds_1 \,,\quad \tilde{\mathbf{A}}_i \equiv \int \bigg( \mathbf{A}_{s_i} - \partial_{s_i} \sum_{j=1}^{i-1} \mathbf{A}_j \bigg) ds_i \,,~~~~ i=2,...,4\,,
\end{equation}
and taking
\begin{equation}
    \tilde{\mathbf{A}}(\epsilon,\vec{s})=\sum_{i=1}^4 \tilde{\mathbf{A}}_i(\epsilon,\vec{s}).
\end{equation}
For this integral family, the matrix $\tilde{\mathbf{A}}(\vec{s})$ is a $\mathbb{Q}$-linear combination of 76 logarithms depending on 11 different square roots. The full set of letters and square roots is presented in Appendix \ref{app:AlphabetPolylogarithmic}. Because the letters contain numerous non-simultaneously rationalizable square roots, it is not manifest that the basis integrals admit polylogarithmic solutions at all orders in $\epsilon$. We will nonetheless refer to the integral sectors composed of the first 71 integrals as the polylogarithmic sectors. This is motivated by the fact that it was shown in Refs. \cite{Bonciani:2016qxi, Bonciani:2019jyb} that the planar Higgs + jet integral families, and the non-planar integral family F, have similar canonical subsectors, for which polylogarithmic results were explicitly obtained at weight 2.

It will also be interesting to study the problem of finding polylogarithmic solutions for these sectors by using the methods recently put forward in \cite{Heller:2019gkq}. 

\subsection{Elliptic integrals}
In the following we will use the notation $B_i{-}B_j$ to denote the range of integrals $B_{i}, B_{i+1}$, $\ldots, B_{j}$. Similarly, we will use the notation $\vec{B}_{i-j}$ to denote the vector $(B_i,B_{i+1},\ldots,B_j)$. 

Integrals $B_{72}{-}B_{84}$ introduce functions of elliptic type. The appearance of functions of elliptic type can be observed at the level of the maximal cut, which gives an indication of the type of functions which appear in the full solution for the integrals in a given sector. When the maximal cut of an integral is elliptic, we expect that the integral cannot be expressed~\cite{Primo:2016ebd, Frellesvig:2017aai, Bosma:2017ens, Harley:2017qut, Primo:2017ipr} in terms of multiple polylogarithms.  Let us discuss the maximal cut for integrals $B_{72}{-}B_{84}$.  These integrals define three integral sectors, i.e. $B_{72}{-}B_{75}$, $B_{76}{-}B_{79}$ and $B_{80}{-}B_{84}$.  Performing the maximal cuts of the basic integral in each of these sectors in $d=4$ we get, using the loop-by-loop Baikov representation~\cite{Frellesvig:2017aai,Harley:2017qut}\footnote{The maximal cut can also be computed by using the loop-by-loop approach in momentum space \cite{Primo:2016ebd}.},
the univariate integrals,
\begin{align}
B_{72}{-}B_{75}: \;\; \text{Cut}( \text{I}_{0111111100} ) &= \int \frac{d z}{s \sqrt{ \big( (z {+} p_4^2)^2 {-} 4 m^2 p_4^2 \big) \, \big( (z {+} t)^2 {+} 4 m^2 t u/s \big)}}, \label{eq:cut126} \\
B_{76}{-}B_{79}: \; \;\text{Cut}(  \text{I}_{1101111100})  & = \int \frac{d z}{s \sqrt{ \big( (z {+} p_4^2)^2 {-} 4 m^2 p_4^2 \big) \, \big( (z {+} u)^2 {+} 4 m^2 t u/s \big)}}, \label{eq:cut123} \\
B_{80}{-}B_{84}: \; \; \text{Cut}( \text{I}_{1111111100} )&= \int \frac{d z}{s \, z \, (z{+}p_4^2{-}s) \, \sqrt{ (z {+} t)^2 {+} 4 m^2 t u/s }}. \label{eq:cut127}
\end{align}
The first two of these evaluate to elliptic integrals of the first kind, while the latter evaluates to a combination of logarithms.
This corresponds to the two elliptic curves,
\begin{align}
y^2 &= \big( (z {+} p_4^2)^2 {-} 4 m^2 p_4^2 \big) \, \big( (z {+} t)^2 {+} 4 m^2 t u/s \big), \nonumber \\
y^2 &= \big( (z {+} p_4^2)^2 {-} 4 m^2 p_4^2 \big) \, \big( (z {+} u)^2 {+} 4 m^2 t u/s \big),
\end{align}
being present in the results.

The integrals $B_{72}{-}B_{75}$ are planar, and indeed that sector is equivalent to the sector of the integral $A_{66}$ discussed in refs~\cite{Bonciani:2016qxi, Francesco:2019yqt}. Likewise the integrals $B_{76}{-}B_{79}$ are merely a crossing thereof with $p_1 \leftrightarrow p_2$, corresponding to $t \leftrightarrow u$. The fact that eq.~\eqref{eq:cut127} does not evaluate to elliptic integrals does not mean that such structures are absent in the un-cut integrals, as elliptic curves would appear at the sub-maximal cuts corresponding to the sectors $B_{72}{-}B_{75}$ and  $B_{76}{-}B_{79}$. 

The appearance of functions of elliptic type can be also observed by analyzing the relevant system of differential equations. Integrals $B_{72}{-}B_{84}$ satisfy,
\begin{equation}
\label{eq:DE elliptic}
\frac{\partial}{\partial s_i}\vec{B}_{72-84}(\vec{s},\epsilon)=\sum_{j=0}^{\infty} \epsilon^j \mathbf{A'}_{s_i}^{(j)}(\vec{s}) \vec{B}_{72-84}(\vec{s},\epsilon)+\vec{G}_{72-84}(\vec{s},\epsilon) ,
\end{equation}
where the vector $\vec{G}_{72-84}(\vec{s},\epsilon)$ depends on the canonical integrals $\vec{B}_{1-71}(\vec{s},\epsilon)$, and the homogeneous matrix has the schematic form,
\begin{equation}
\label{eq:EllipticDE}
\mathbf{A'}_{s_i}^{(0)}(\vec{s})=\left(
\begin{array}{cccc|cccc|ccccc}
 \bm{*} & \;0\; & \;0\; & \bm{*} & \;0\; & \;0\; & \;0\; & \;0\; & \;0\; & \;0\; & \;0\; & \;0\; & \;0\; \\
 \bm{*} & 0 & 0 & \bm{*} & 0 & 0 & 0 & 0 & 0 & 0 & 0 & 0 & 0 \\
 \bm{*} & 0 & \bm{*} & \bm{*} & 0 & 0 & 0 & 0 & 0 & 0 & 0 & 0 & 0 \\
 \bm{*} & 0 & 0 & \bm{*} & 0 & 0 & 0 & 0 & 0 & 0 & 0 & 0 & 0 \\
 \hline
 0 & 0 & 0 & 0 & \bm{*} & 0 & 0 & \bm{*} & 0 & 0 & 0 & 0 & 0 \\
 0 & 0 & 0 & 0 & \bm{*} & 0 & 0 & \bm{*} & 0 & 0 & 0 & 0 & 0 \\
 0 & 0 & 0 & 0 & \bm{*} & 0 & \bm{*} & \bm{*} & 0 & 0 & 0 & 0 & 0 \\
 0 & 0 & 0 & 0 & \bm{*} & 0 & 0 & \bm{*} & 0 & 0 & 0 & 0 & 0 \\
 \hline
 \bm{*} & 0 & 0 & \bm{*} & \bm{*} & 0 & 0 & \bm{*} & 0 & 0 & 0 & 0 & 0 \\
 \bm{*} & 0 & 0 & \bm{*} & \bm{*} & 0 & 0 & \bm{*} & 0 & 0 & 0 & 0 & 0 \\
 \bm{*} & 0 & 0 & \bm{*} & \bm{*} & 0 & 0 & \bm{*} & 0 & 0 & 0 & 0 & 0 \\
 \bm{*} & 0 & 0 & \bm{*} & \bm{*} & 0 & 0 & \bm{*} & 0 & 0 & 0 & 0 & 0 \\
 \bm{*} & 0 & 0 & \bm{*} & \bm{*} & 0 & 0 & \bm{*} & 0 & 0 & 0 & 0 & 0 \\
\end{array}
\right),
\end{equation}
where the lines separate the 3 elliptic sectors. In \cite{Primo:2016ebd} it was observed that the homogeneous differential equation of a given integral is solved by the maximally cut integral. This implies that, when the maximal cut is elliptic, the solution of the integral can be expressed in terms of iterated integrals over functions of elliptic type. As seen from Eqs. (\ref{eq:cut126}) and (\ref{eq:cut123}), we encounter this scenario for integrals $B_{72}{-}B_{75}$ and $B_{76}{-}B_{79}$. On the other hand, the maximal cut of sector $B_{80}{-}B_{84}$ is logarithmic. However, as seen from Eq.~(\ref{eq:EllipticDE}), this sector couples to the lower elliptic sectors via inhomogeneous terms of the differential equations, implying that these integrals can be expressed in terms of iterated integrals over the same functions of elliptic type.

We remark that the presence of multiple elliptic curves renders the functional form of the solution an open problem (but see e.g. \cite{Adams:2018bsn} for progress in this direction).

%The homogeneous part of the differential equations for the integrals in a given sector is solved by the maximal cut \cite{Primo:2016ebd}. In our case, the homogeneous solutions of integrals 72 and 76 are linear combinations of the elliptic integrals (\ref{eq:cut126}) and (\ref{eq:cut123}) respectively, evaluated along all possible pairs of branching points of the respective elliptic curve \cite{Bosma:2017ens,Primo:2017ipr,Harley:2017qut}. By using the variation of the constants method, it is possible to solve integrals 72 and 76 in terms of iterated integrals over functions of elliptic type (the integration kernels depend on products of the homogeneous solutions). The remaining integrals 73-75, 77-84 are coupled to 72 and 76 via inhomogeneous terms of the differential equations, and they can be solved in terms of the same class of functions, i.e. iterated integrals over functions of elliptic type (the elliptic integrals appearing in the integration kernels are the homogeneous solutions of 72,76).

\section{Series expansion along contours}
\label{sec:series_exp}
We consider the series expansion strategy outlined in \cite{Bonciani:2019jyb,Francesco:2019yqt} (see also \cite{Pozzorini:2005ff,Aglietti:2007as,Mueller:2015lrx,Lee:2017qql,Lee:2018ojn,Bonciani:2018uvv,Mistlberger:2018etf} for the application of expansion methods to single scale processes, and ~\cite{Melnikov:2016qoc,Melnikov:2017pgf,Bonciani:2018omm, Bruser:2018jnc,Davies:2018ood,Davies:2018qvx} for expansion methods applied to multiscale problems in particular kinematic limits). The strategy relies on parametrizing the integrals along straight line segments, for which we solve the corresponding differential equations in terms of one-dimensional generalized series. We briefly review the strategy here, and highlight aspects that are specific to the integral family under consideration. We start from the system of differential equations of the basis defined in Appendix \ref{app:CanonicalBasis}, which has the form,
\begin{align}
    d\vec{B} = \mathbf{M} \vec{B}\,,
\end{align}
where $\mathbf{M}=\sum_{s_i}\mathbf{M}_{s_i}(\epsilon, \vec{s})ds_i$, and where we otherwise suppress variable dependence in the notation. For convenience, we put $m^2=1$ without loss of generality. We consider a generic line parametrized as,
\begin{equation}
\vec{\gamma}(\lambda) = \{\gamma_s(\lambda),\gamma_t(\lambda),\gamma_{p_4^2}(\lambda)\}\,.
\end{equation}
The differential equations along this line take the form,
\begin{align}
    \frac{\partial}{\partial \lambda} \vec{B} = \mathbf{M}_\lambda \vec{B}\,,
\end{align}
where $\mathbf{M}_\lambda = \sum_{s_i}\mathbf{M}_{s_i}(\epsilon, \vec{s})\frac{\partial \gamma_{s_i}}{\partial \lambda}$. Next, we expand the differential equations in $\epsilon$ to obtain a system for each order in $\epsilon$. In particular, we let,
\begin{align}
    \vec{B} = \sum_{k=0}^\infty B^{(k)} \epsilon^k,\quad \mathbf{M} = \sum_{k=0}^\infty \mathbf{M}^{(k)} \epsilon^k \,.
\end{align}
Note that both expansions start at finite order for our choice of basis. %\footnote{We remark that the matrix $\mathbf{M}_{m^2}$ does have a component $\epsilon^{-1}$ for our basis, but we can ignore it since we work in the kinematics $m^2 = 1$.} 
The system of differential equations now takes the following form, order-by-order in $\epsilon$,
\begin{align}
    \frac{\partial}{\partial \lambda}\vec{B}^{(i)} = \mathbf{M}^{(0)}_\lambda \vec{B}^{(i)} + \sum_{k=1}^i \mathbf{M}_\lambda ^{(k)} \vec{B}^{(i-k)}\,,
\end{align}
where we separated out the homogeneous part $\mathbf{M}^{(0)}_\lambda \vec{B}^{(i)}$ from the inhomogeneous part. The homogeneous matrix $\mathbf{M}^{(0)}$ determines the sequence in which the individual integrals should be integrated, and which integrals are coupled.

Let us consider first the polylogarithmic sectors and review the series solution strategy for those. The system of differential equations becomes simply,
\begin{align}
    \frac{\partial}{\partial \lambda}\vec{B}^{(i)} = \mathbf{M}_\lambda ^{(1)} \vec{B}^{(i-1)}\,,
\end{align}
where $\mathbf{M}_\lambda^{(1)}$ = $\epsilon \left(\partial \tilde{\mathbf{A}}/\partial \lambda\right)$, and $\tilde{\mathbf{A}}$ was defined in Eq. (\ref{eq:d_canonical}). Hence, the work for the polylogarithmic sectors amounts to solving a sequence of first order differential equations without homogeneous parts. The general solution is easily found from a single integration,
\begin{align}
    \label{eq:polylogintegration}
    \vec{B}^{(i)} (\lambda) = \int_{\lambda_0}^{\lambda_1} \mathbf{M}_{\lambda} ^{(1)} \vec{B}^{(i-1)} d\lambda + \vec{B}^{(i)} (\lambda_0)\,.
\end{align}
Importantly, we solve the integration by performing series expansions in $\lambda$. The only algebraic terms in our basis and in the matrices are square roots, so that the expansions of the matrix elements are in terms of integer and half-integer powers of $\lambda$. After integrating, the series expansions will also contain logarithmic terms. Therefore, in general each integration in Eq. (\ref{eq:polylogintegration}) is of the form,
\begin{align}
    \int \lambda^q \log(\lambda)^p\,,
\end{align}
where $q$ is an integer or half-integer, and $p$ is a non-negative integer. It may easily be verified that such integrals evaluate to sums of terms of the same type, by using integration-by-parts identities to reduce the power of the logarithm inside the integral.

As shown in Eq. (\ref{eq:EllipticDE}), we simplified our basis in such a way that in each elliptic sector at most 2 integrals are coupled together, specifically the pairs of integrals $B_{72}$, $B_{75}$ and $B_{76}$, $B_{79}$. These integrals can be solved by combining their first order differential equations into second order differential equations. Integrals $B_{74}$ and $B_{78}$ can be solved from their first order differential equation, but, in contrast to the polylogarithmic sectors, their differential equations have a homogeneous component. Lastly, integrals $B_{73}$, $B_{77}$, $B_{80}$, $B_{81}$, and $B_{82}$ satisfy a first order differential equation without a homogeneous component, and can be solved in the same manner as the polylogarithmic integrals. For completeness, we discuss solving these two cases next.

Consider a first order differential equation with homogeneous component of the form,
\begin{align}
    \label{eq:firstorderwithom}
    f'(\lambda) + a(\lambda)f(\lambda) + b(\lambda) = 0\,.
\end{align}
The solution to the homogeneous part is easily found to be,
\begin{align}
    \mu(\lambda) = e^{-\int a(\lambda) d\lambda}\,,
\end{align}
up to an arbitrary multiplicative constant. The full solution to Eq. (\ref{eq:firstorderwithom}) is then given in terms of $\mu(\lambda)$ by,
\begin{align}
    f(\lambda) = \mu(\lambda)\left[-\int\frac{ b(\lambda)}{\mu(\lambda)}d\lambda+c\right]\,.
\end{align}
%Note that freedom to rescale $\mu(\lambda)$ cancels out in the full solution.

Now, consider a second order differential equation of the form,
\begin{align}
    \label{eq:2ndordergeneral}
    f''(\lambda) + a(\lambda)f'(\lambda) + b(\lambda)f(\lambda) + c(\lambda) = 0\,.
\end{align}
Given two solutions $\mu_1(\lambda)$ and $\mu_2(\lambda)$ to the homogeneous part of the differential equation, the general solution can be written using the method of variation of parameters as
\begin{align}
    \label{eq:variationofparameters2ndorder}
    f(\lambda) &=  \mu_1(\lambda) \int \frac{\mu_2(\lambda) c(\lambda)} {\mu_1(\lambda)\mu_2'(\lambda)-\mu_2(\lambda)\mu_1'(\lambda)}d\lambda\ -  \mu_2(\lambda) \int \frac{\mu_1(\lambda) c(\lambda) }{\mu_1(\lambda)\mu_2'(\lambda)-\mu_2(\lambda)\mu_1'(\lambda)}d\lambda  \nonumber\\
    &+ d_1 \mu_1(\lambda) + d_2 \mu_2(\lambda)\,,
\end{align}
where $d_1$ and $d_2$ are complex constants to be fixed from boundary conditions. The remaining challenge is to find two distinct homogeneous solutions $\mu_1(\lambda)$ and $\mu_2(\lambda)$ that are not related by a rescaling. From the well-known Frobenius method (see e.g.~\cite{Coddington} for an extensive review of the method), it follows that we may always find one series solution of the form $\mu_1(\lambda) = \lambda^r +  \lambda^r\sum_{k=1}^\infty \mu_{1,k} \lambda^k$. The values for $r$ and $\mu_{1,k}$ may be found up to the desired order in $\lambda$ by plugging $\mu_1(\lambda)$ into the homogeneous differential equation as an ansatz, and solving order-by-order in $\lambda$ for the unknowns. The lowest order in $\lambda$ gives a quadratic equation in $r$ called the indicial equation. By picking $r$ to be the largest root of the indicial equation, it is guaranteed that we may solve for the remaining unknowns $\mu_{1,k}$ with $k\geq 1$.

It remains to find a second homogeneous solution. This may be done in the following way. First we write the second homogeneous solution $\mu_2(\lambda)$ as $\mu_2(\lambda) = \mu_1(\lambda)h(\lambda)$. Plugging this expression into the homogeneous part of Eq. (\ref{eq:2ndordergeneral}), we find a new equation,
\begin{align}
    \mu _1(\lambda ) h''(\lambda )+h'(\lambda ) \left(a(\lambda ) \mu _1(\lambda )+2 \mu _1'(\lambda )\right)=0\,,
\end{align}
which we recognize as a first order homogeneous differential equation for $h'(\lambda)$, which we know how to solve. This way, we obtain the second homogeneous solution $\mu_2(\lambda)$. Thus we may now use Eq. (\ref{eq:variationofparameters2ndorder}) to compute the full solution to Eq. (\ref{eq:2ndordergeneral}).

\subsection{Boundary conditions}

To fix our system of differential equations we need a suitable boundary point. Similar to \cite{Bonciani:2019jyb}, we work in the heavy mass limit parametrized by,
\begin{align}
    \gamma_{hm}(\lambda) = \{s \lambda,t \lambda, p_4^2 \lambda\}\,,
\end{align}
where $\lambda$ is a line parameter that goes to zero. Using the method of asymptotic expansions in the parametric representation \cite{Beneke:1997zp,Jantzen:2011nz,Semenova:2018cwy,Smirnov:1999bza,Pak:2010pt, Jantzen:2012mw}, we may obtain values of our basis integrals in the heavy mass limit. The final result turns out to be very simple,
\begin{align}
    \label{eq:heavymasslimitexplicit}
    \lim_{\lambda \rightarrow 0} B_1(\gamma_{hm}(\lambda)) = e^{2 \gamma_E  \epsilon } \Gamma (1+\epsilon)^2 (m^2)^{-2 \epsilon } \,,\quad \lim_{\lambda \rightarrow 0} B_i(\gamma_{hm}(\lambda)) = 0 \quad \text{ for }i=2,\ldots,84\,.
\end{align}
We note that the homogeneous solution of the differential equation for $B_{78}$ along $\gamma_{hm}(\lambda)$ is proportional to $\lambda$, and hence we are not able to determine the boundary constant for $B_{78}$ directly from Eq. (\ref{eq:heavymasslimitexplicit}). It may be verified that $B_{78}$ is also zero at order $\lambda^1$ in the heavy mass limit, and hence the constant multiplying the homogeneous solution may be put to zero for this integral.

\subsection{Convergence of the series}

\begin{figure}
  \centering
  \begin{tabular}{@{}c@{}}
    \includegraphics[width=\linewidth]{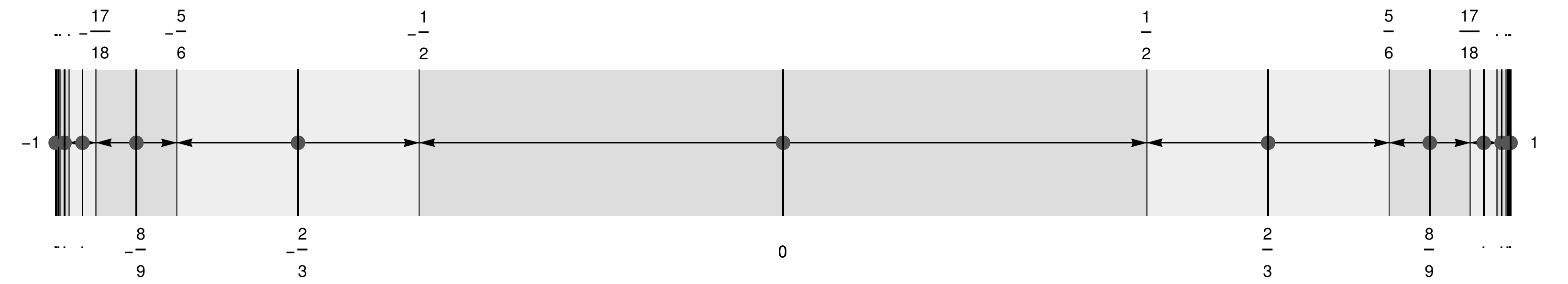} \\[\abovecaptionskip]
    \small (a)
  \end{tabular}

  \vspace{\floatsep}

  \begin{tabular}{@{}c@{}}
    \includegraphics[width=\linewidth]{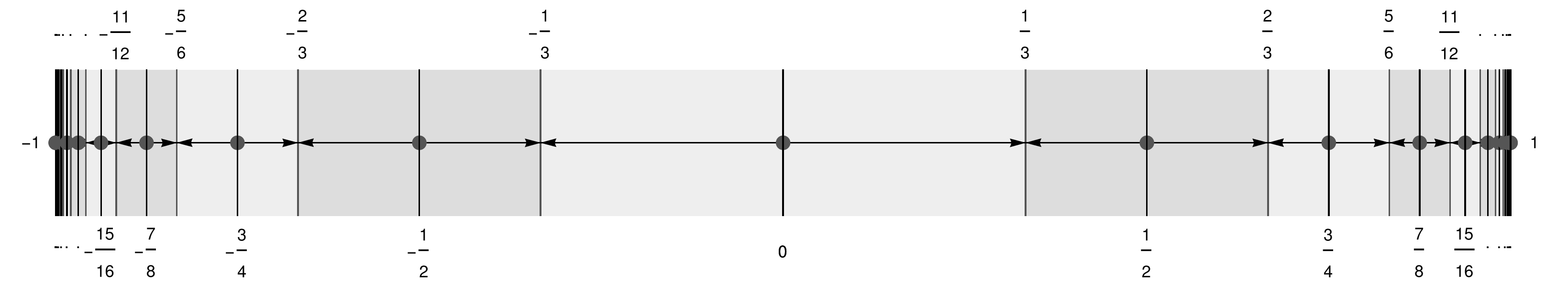} \\[\abovecaptionskip]
    \small (b) 
  \end{tabular}

  \caption{These figures illustrate subdivisions of an expansion in the unit interval [-1,1] with singularities at $-1, 0$ and $1$, in terms of additional expansions, such that each expansion can be matched to the next one at a fixed fraction of the distance to its nearest singularities. The numbers on top are the matching points between neighbouring expansions, while the numbers at the bottom indicate the expansions points for (a) $k = 2$: Moving at most half the distance to the nearest singularity,  (b) $k = 3$: Moving at most one-third the distance to the nearest singularity.}
  \label{fig:figmobiusrefine}
\end{figure}

A trait of the expansion strategy is that each expansion at a given point along a line has a limited range of convergence. Namely, each expansion at a given point is valid up to the distance of the point to the nearest singularity. Thus, to obtain results along a given line, numerous expansions along segments of the line have to be patched together in order to reach a given point in phase space. In particular, to cross a singularity we may perform an expansion at the singularity, and fix its boundary conditions from an expansion at a neighbouring point along the line.  We employ the following strategy for deciding along which line segments to expand:
\begin{itemize}
  \item First we create a list $A$ of all singularities of the matrix elements of $\mathbf{M}_\lambda$ on the line $\gamma(\lambda)$ along which we seek to integrate. By singular point we mean any non-analytic point of the differential equations. In our case, these are the zeros of the denominators of the matrix elements, and the zeros of the square roots.
  \item Some of the singularities may be complex. We replace each complex singularity $\lambda^{\text{sing}} = \lambda^{\text{sing}}_{re} + i \lambda^{\text{sing}}_{im}$ in the list $A$ by three real points: $\lambda^{\text{sing}}_{re} - \lambda^{\text{sing}}_{im}$, $\lambda^{\text{sing}}_{re}$ and $\lambda^{\text{sing}}_{re} + \lambda^{\text{sing}}_{im}$\,.
  \item Next, we consider a M{\"o}bius transformation $\lambda=g(\lambda')$ for each triplet $(a,b,c)$ of neighbouring points in $A$, such that $g^{-1}(\{a,b,c\}) = \{-1,0,1\}$. Note that a series (in $\lambda'$) centered at $\lambda' = 0$ will have a radius of convergence greater than or equal to 1.
  \item To obtain results along $\gamma(\lambda)$ from $\lambda_0$ to $\lambda_1$, we have to match expansions along neighbouring line segments, which are expressed in terms of M{\"o}bius transformed line parameters, say $\lambda'$ and $\lambda''$. We may find a matching point between two neighbouring expansions by solving $\lambda'=-\lambda''$, assuming $\lambda''$ corresponds to the line segments lying on the right.
  \item In general, one may find that this condition picks $\lambda'$ and $\lambda''$ to be very close to 1 and -1, respectively, where both series may be very slowly converging. This can be solved by adding additional expansion points along the line segments. In particular, we may consider new expansion points between -1 and 1, such that upon matching neighbouring expansions, neither gets evaluated further than a certain fraction of the distance to the nearest singularity. We will refer to the inverse of this fraction by the parameter $k$. For example, with $k$ = 2, the expansion points are chosen such that no series is evaluated beyond half its radius of convergence. The situation is illustrated in Figure \ref{fig:figmobiusrefine}. By choosing higher values of $k$, we will increase the precision of the results, since the expansions along each line segment are evaluated closer to the origin.
\end{itemize}
We note that in general we may encounter both spurious, physical, and non-physical singularities. The spurious singularities are singularities that only appear in the elements of $\mathbf{M}$, but which are not singularities of the basis integrals themselves. The physical singularities are threshold singularities, in our case $s=4m^2$ and $p_4^2 = 4m^2$. For those, it is important to cross the singularity according to Feynman prescription, which tells us to interpret $s$ and $p_4^2$ as having an infinitesimally small positive imaginary part. Furthermore, we should make sure to assign the same imaginary part to the square roots in our basis that are associated with physical singularities. Specifically, some of our basis integrals have the prefactors $\sqrt{4m^2-p_4^2}$ and $\sqrt{4m^2-s}$, which are analytically continued as $\sqrt{4m^2-p_4^2-i\delta}$ and $\sqrt{4m^2-s-i\delta}$ for an infinitesimally small $\delta>0$. Lastly, there are also non-physical singularities, which can arise from rational prefactors in the basis, or from square roots in the basis that do not correspond to physical singularities. Since these singularities are introduced by the basis choice, we are free to assign every non-physical root in the basis the standard branch, i.e. we consider the argument to carry the imaginary part $+i\delta$.

To improve the convergence of our series solutions, we compute their diagonal Padé approximants and evaluate those instead at each (matching) point. Since we are dealing with generalized series that may in general include powers of logarithms, we collect first on powers of logarithms and compute the Padé approximant for each series that multiplies a given power.

\begin{figure}
    \centering
    \includegraphics[width=0.6\linewidth]{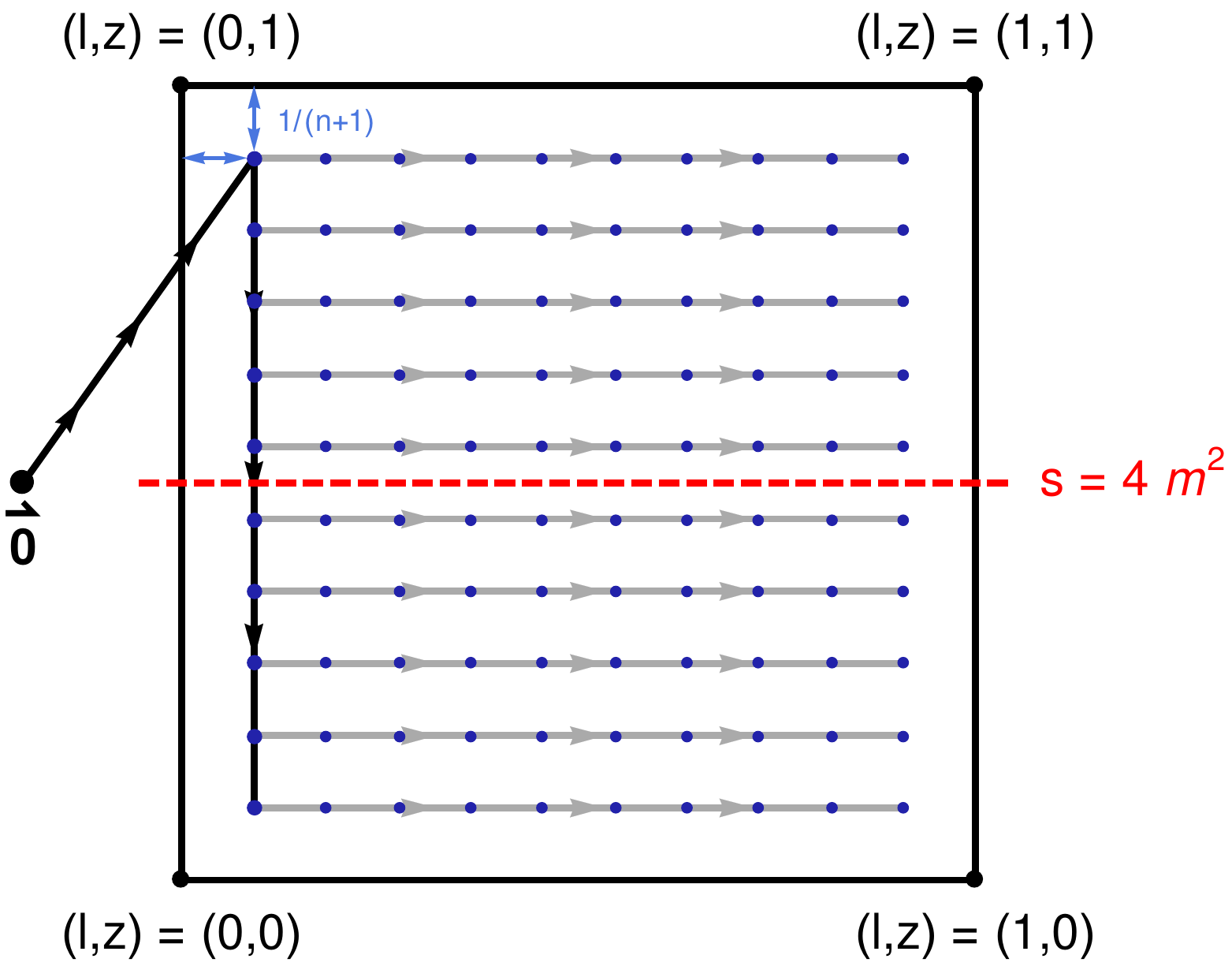}
    \caption{Depiction of lines along which we produce samples in the physical region of the top. The black lines {\footnotesize $\vec{0}\rightarrow \left(1/(n+1), n/(n+1)\right) \rightarrow \left(1/(n+1), 1/(n+1)\right)$} are computed first to obtain boundary values for $n$ horizontal lines, depicted in grey. The horizontal lines are themselves used to produce $n$ evenly spaced samples, denoted by blue dots. The particle production threshold $s = 4m^2$ is depicted by a dashed red line. Depicted is the case with $n = 10$. The actual plots are produced with $n = 100$.}
    \label{fig:topregion}
\end{figure}

\section{Results for top and bottom quarks}
\label{sec:results}

\begin{figure}[ht]
	\begin{tabular}{ p{3.5cm}  p{3.5cm}  p{3.5cm}  p{3.5cm}}
        \imgBotRe{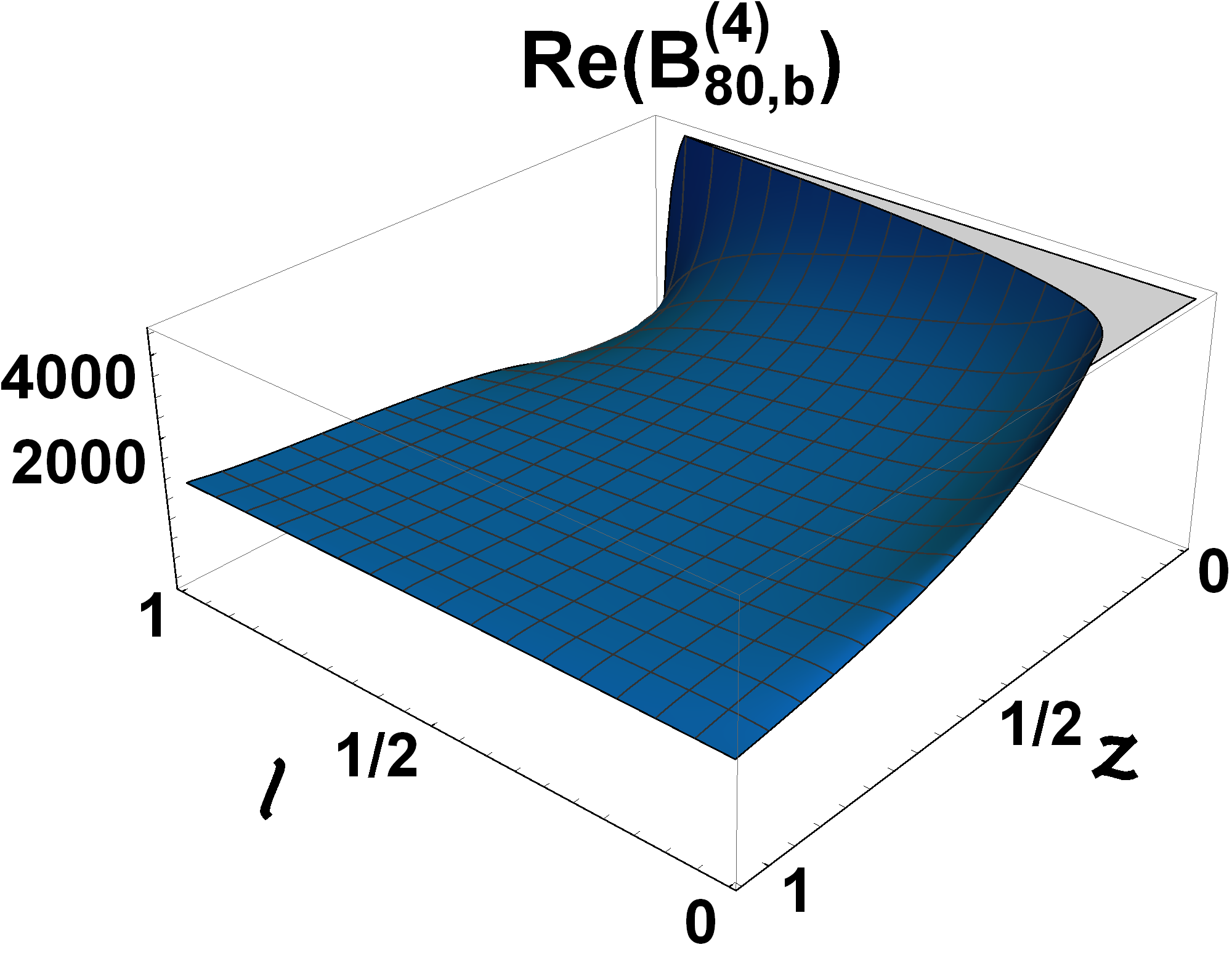} & \imgBotIm{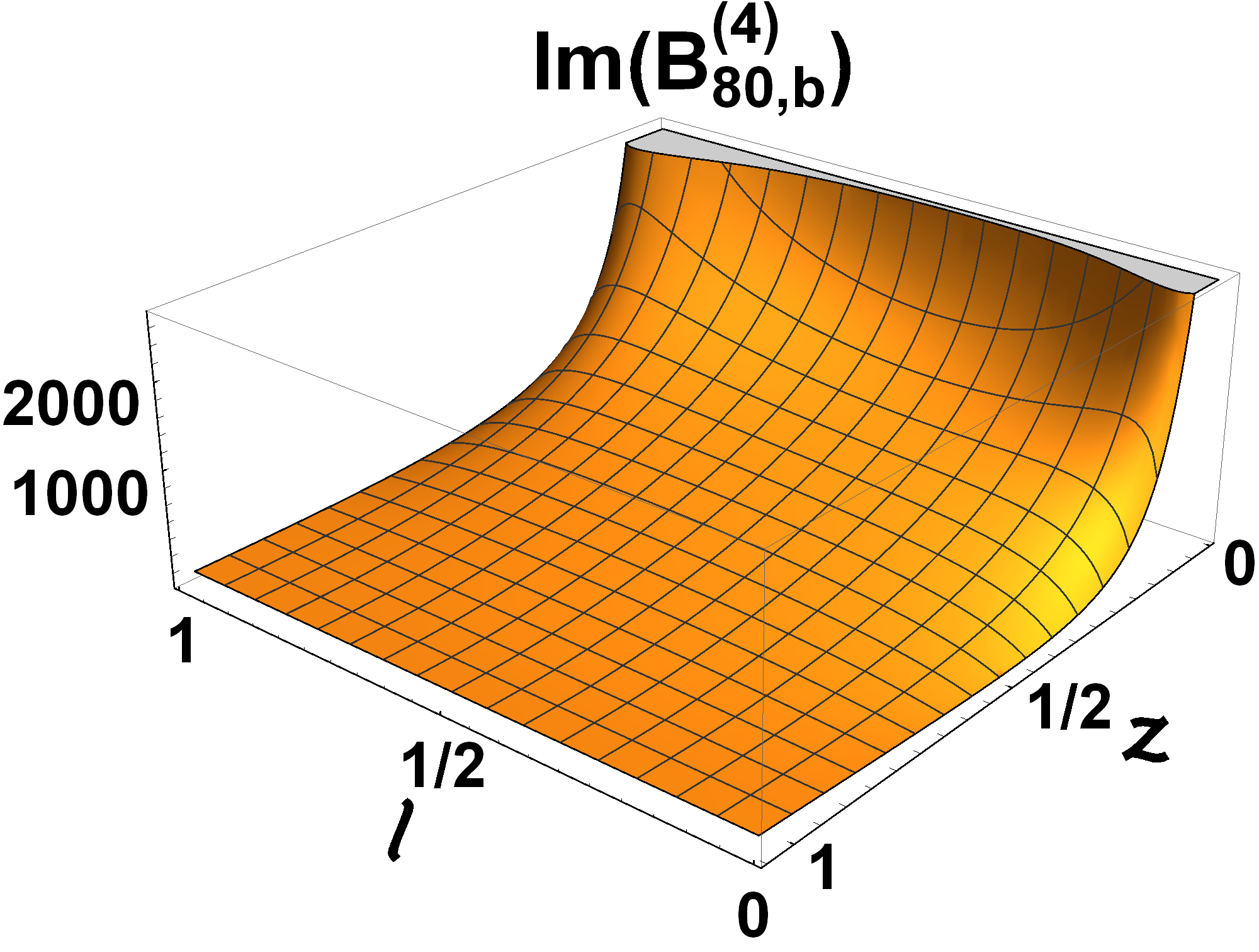} & \imgTopRe{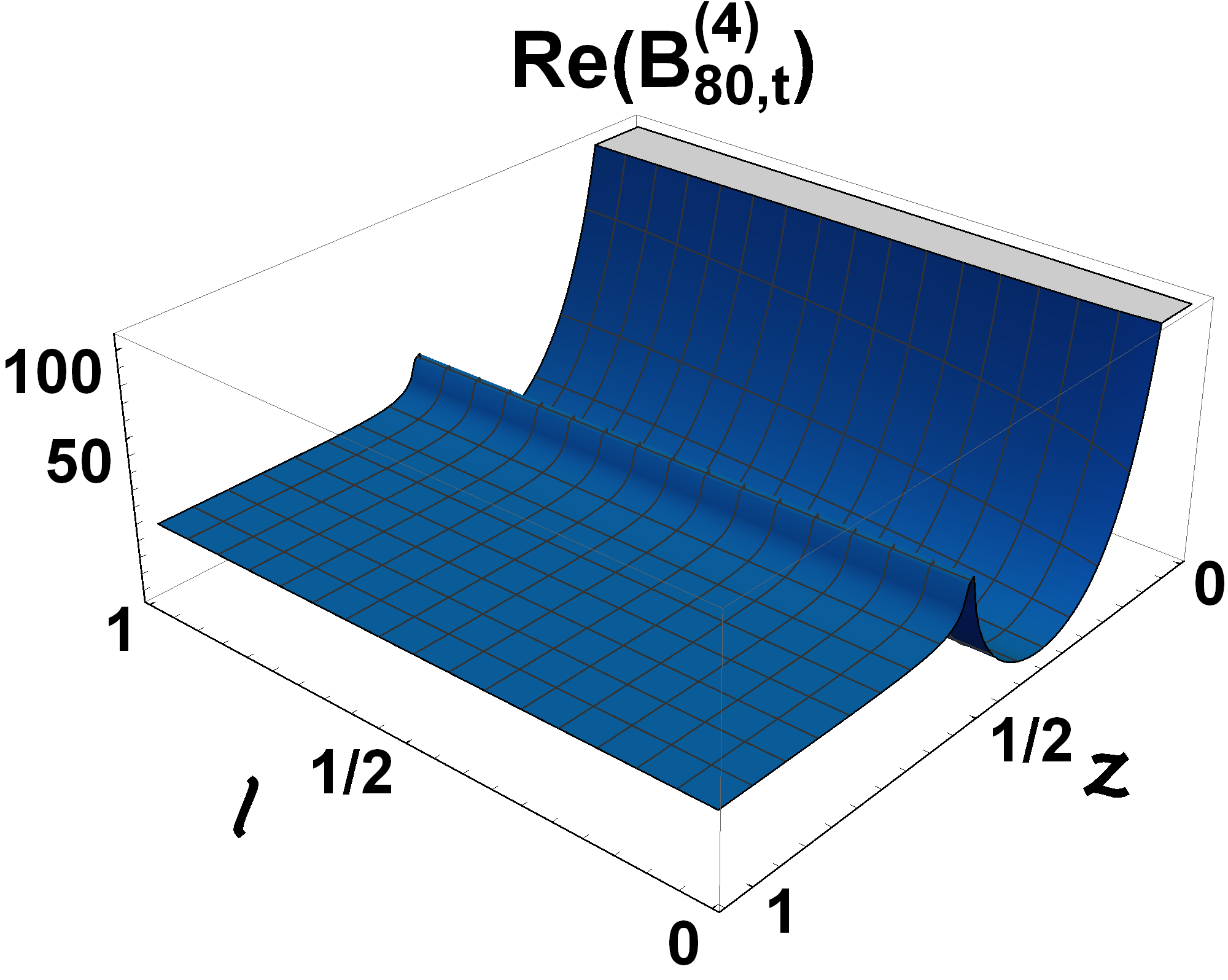} & \imgTopIm{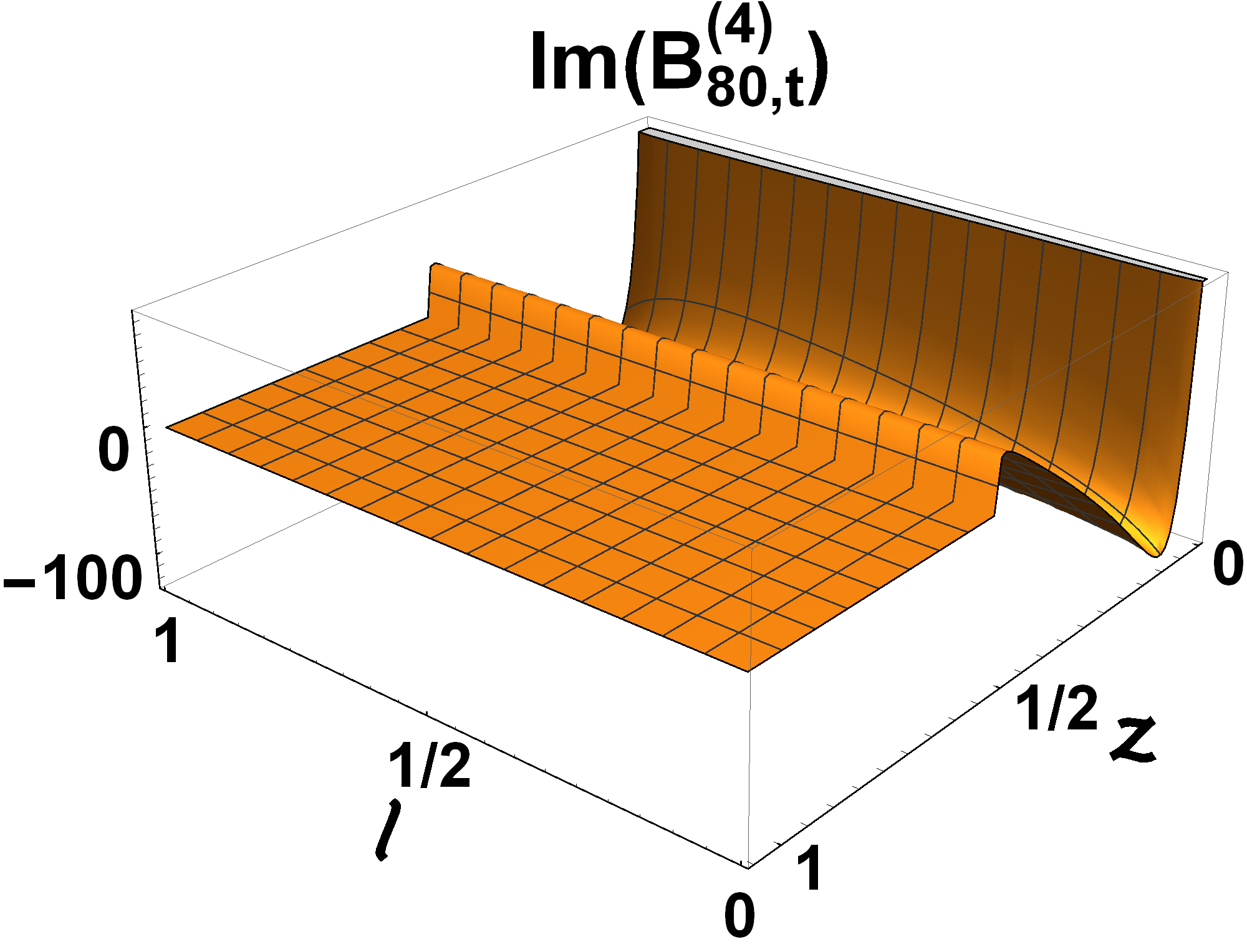} \\
        \imgBotRe{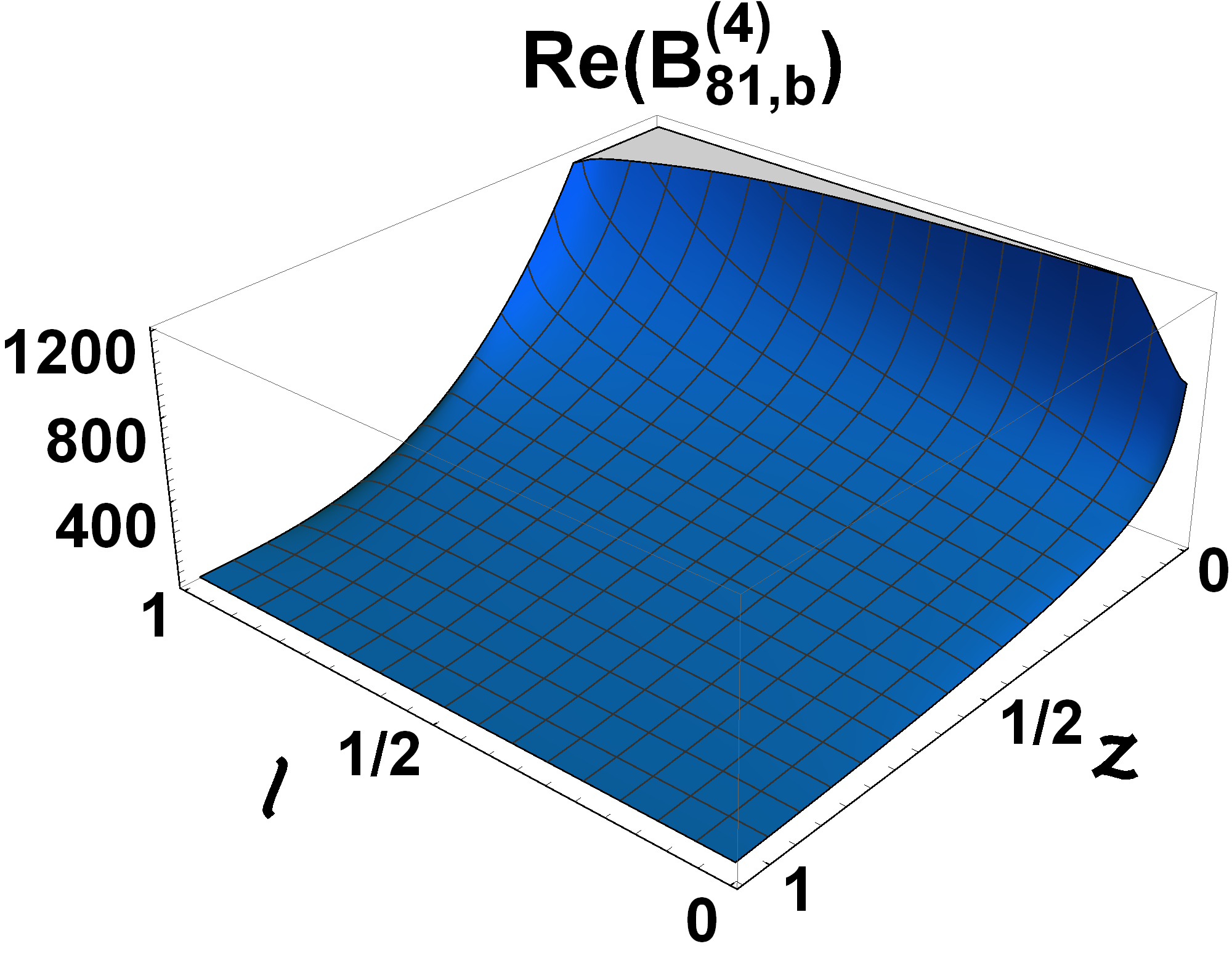} & \imgBotIm{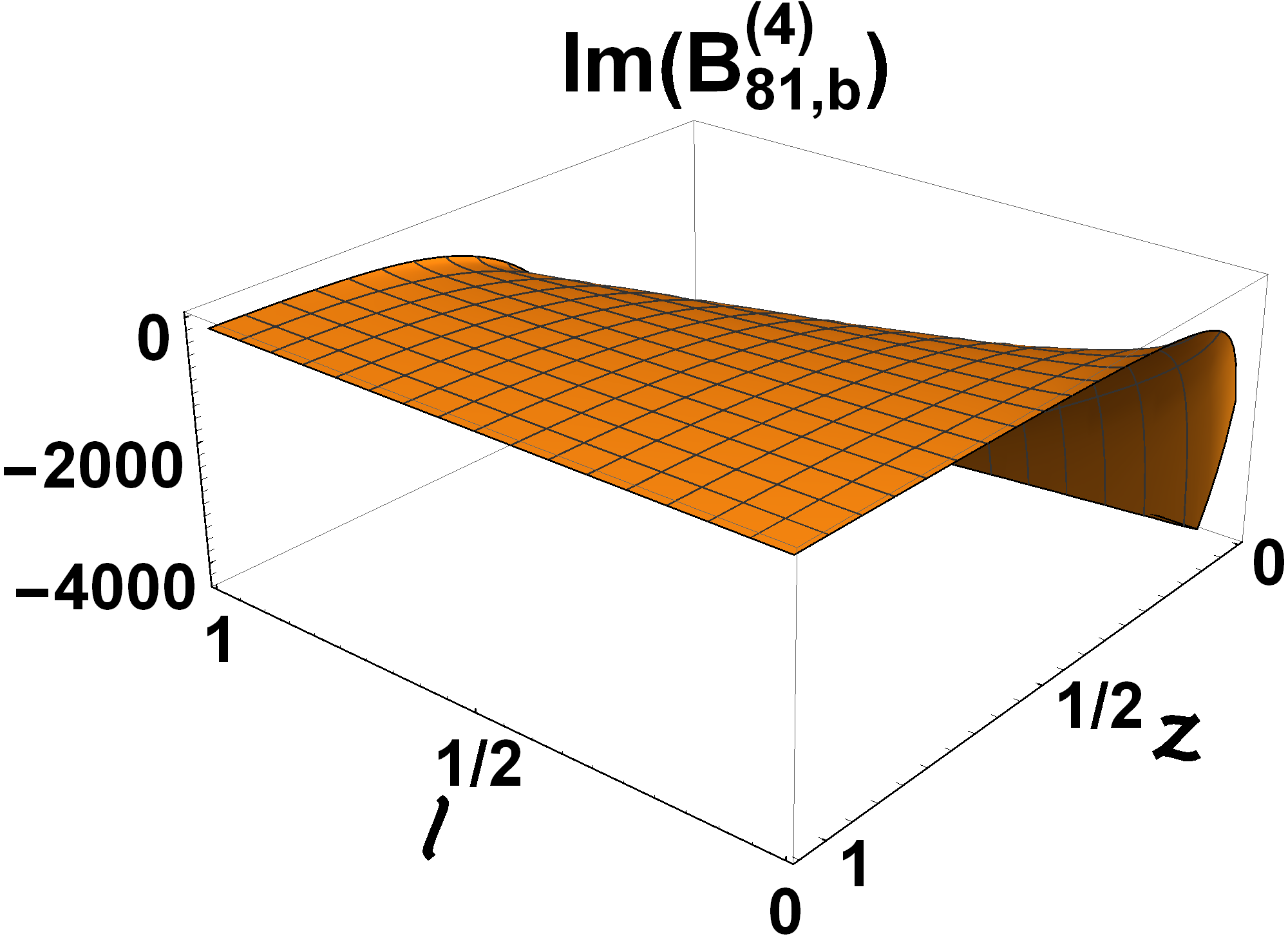} & \imgTopRe{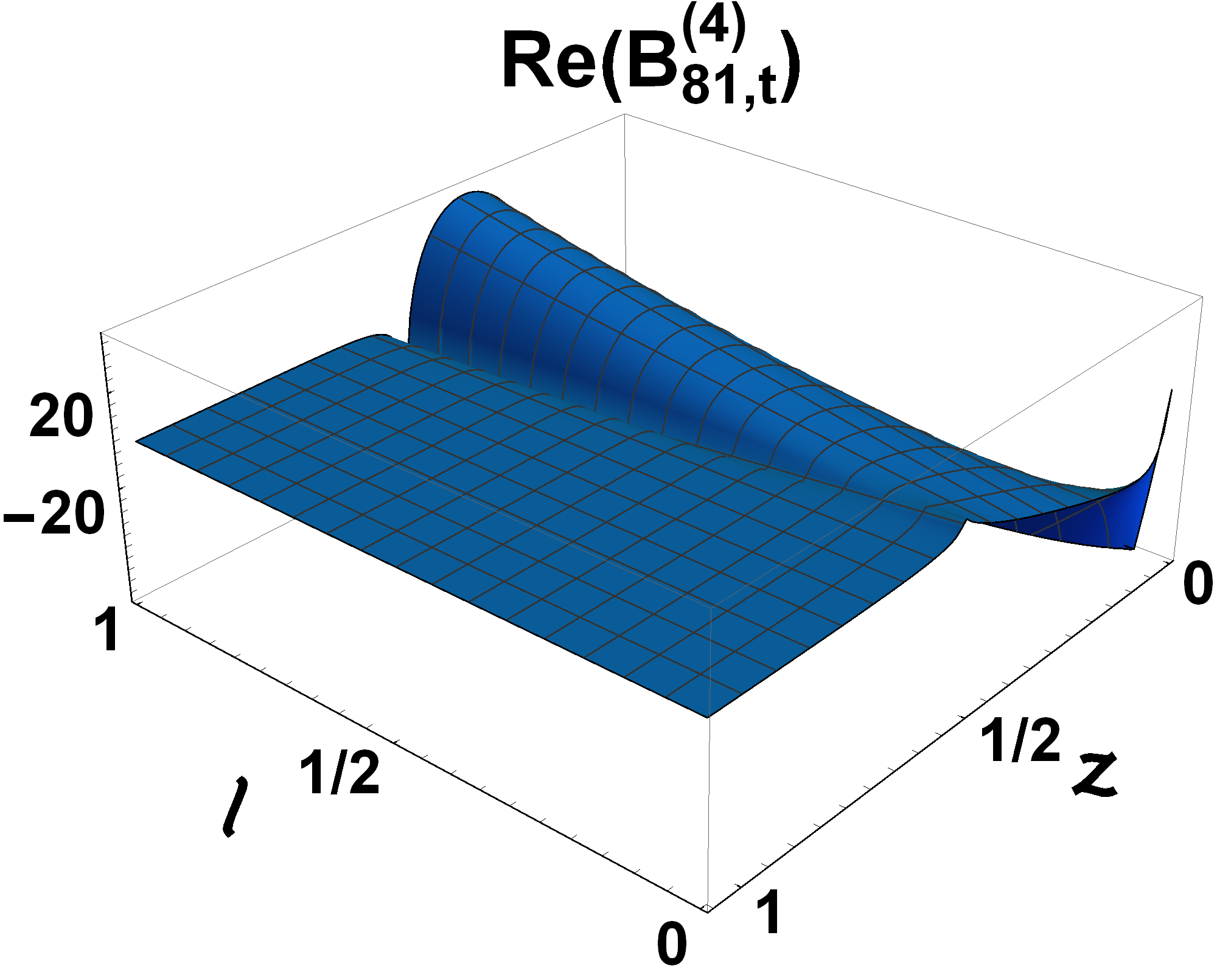} & \imgTopIm{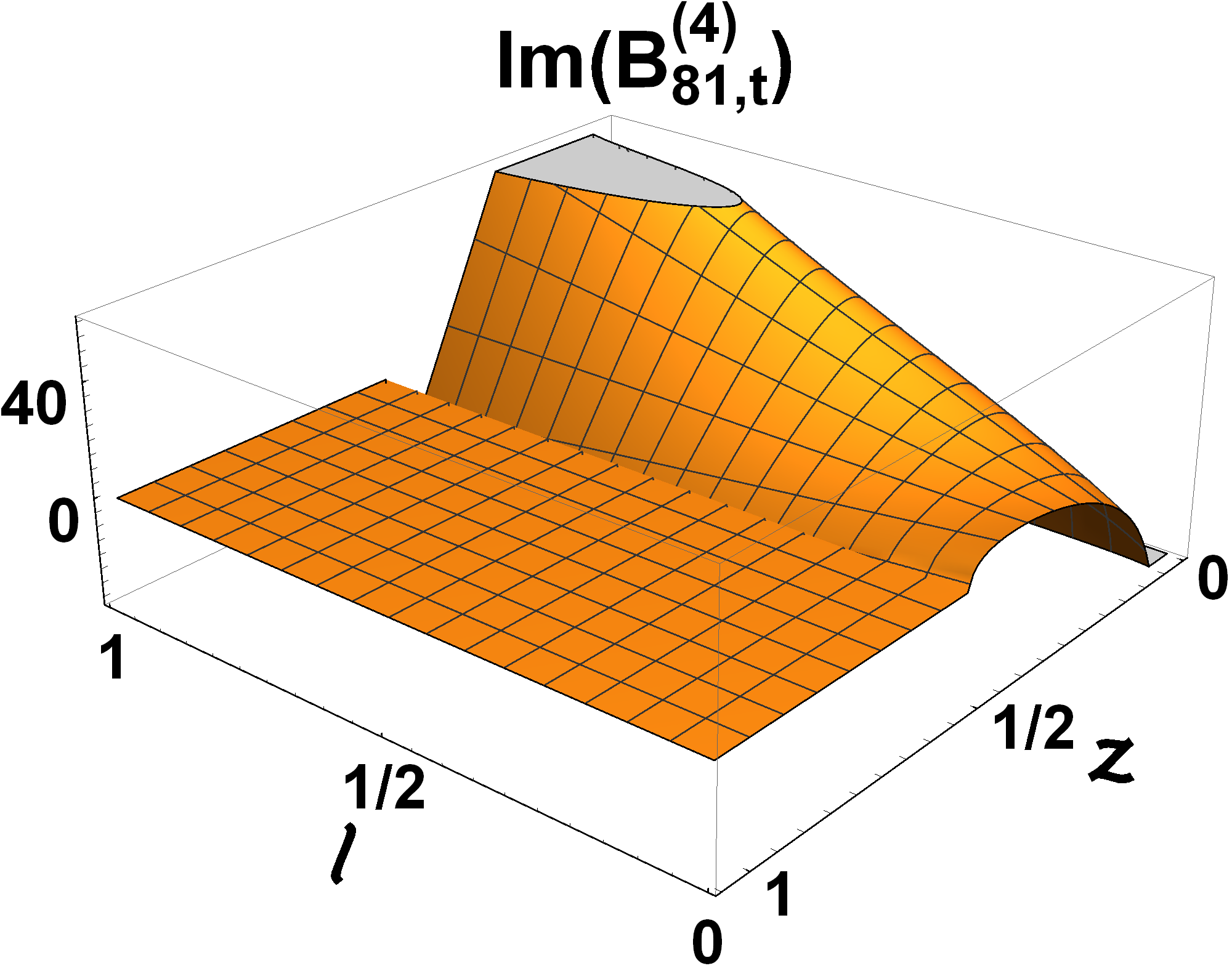} \\
        \imgBotRe{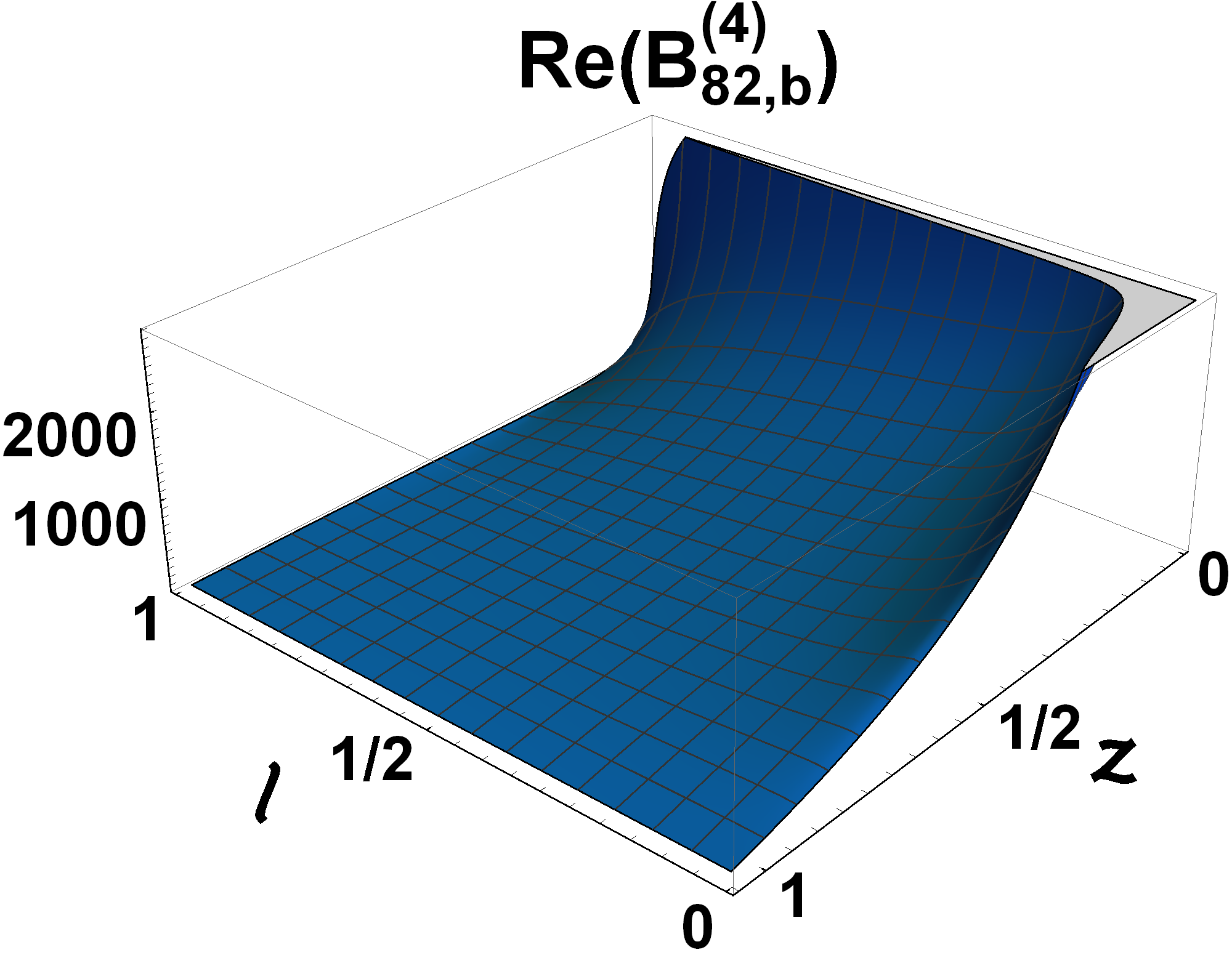} & \imgBotIm{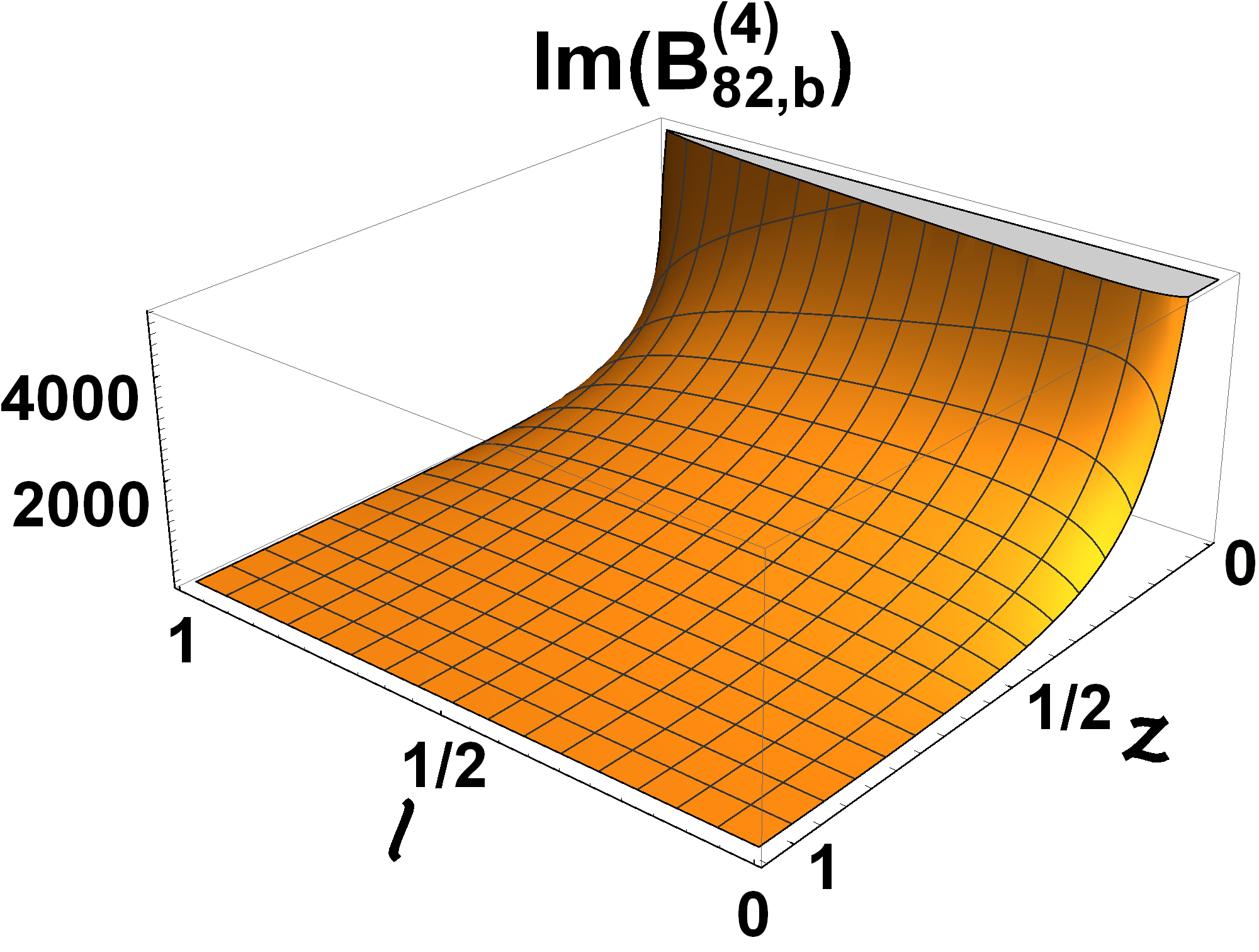} & \imgTopRe{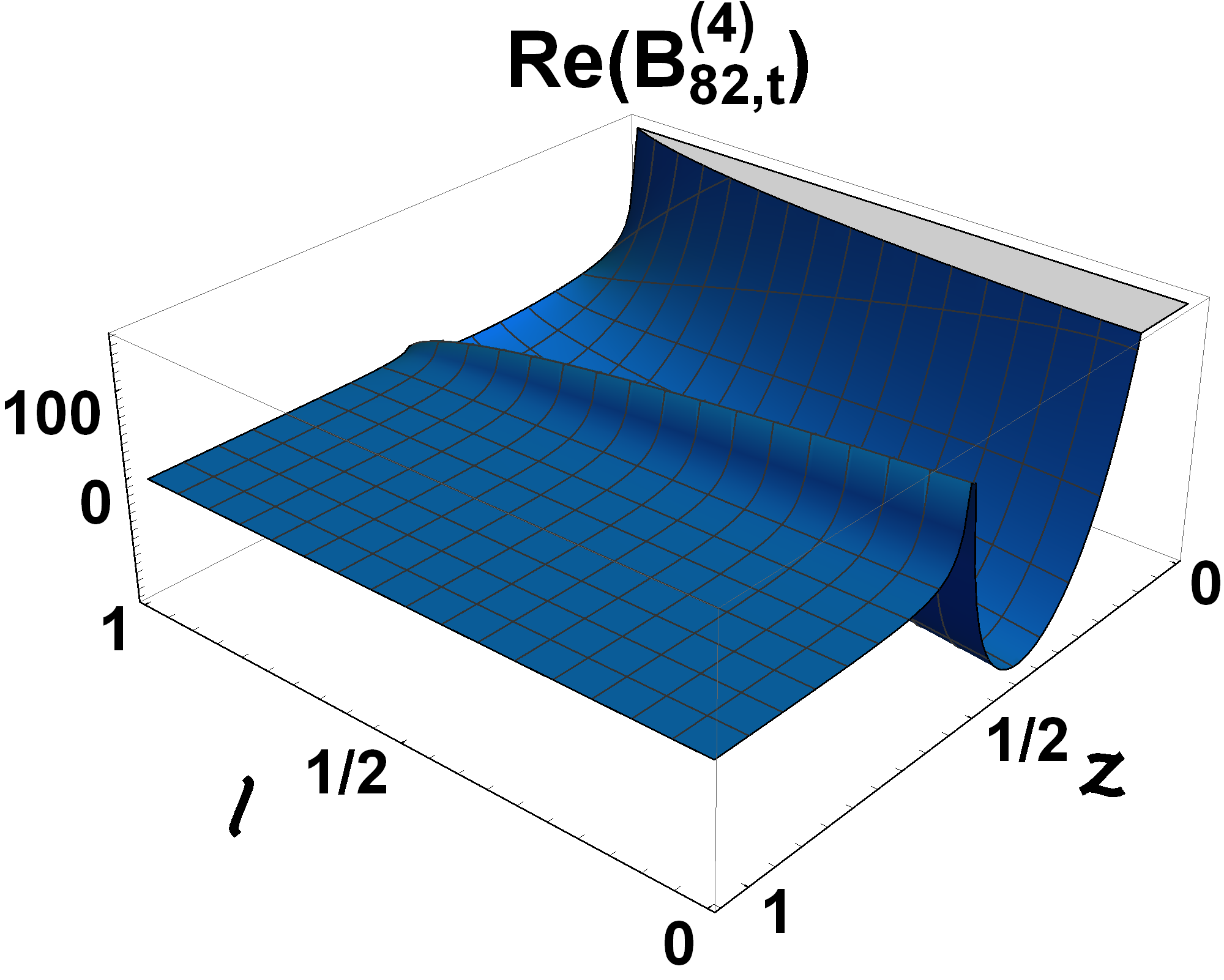} & \imgTopIm{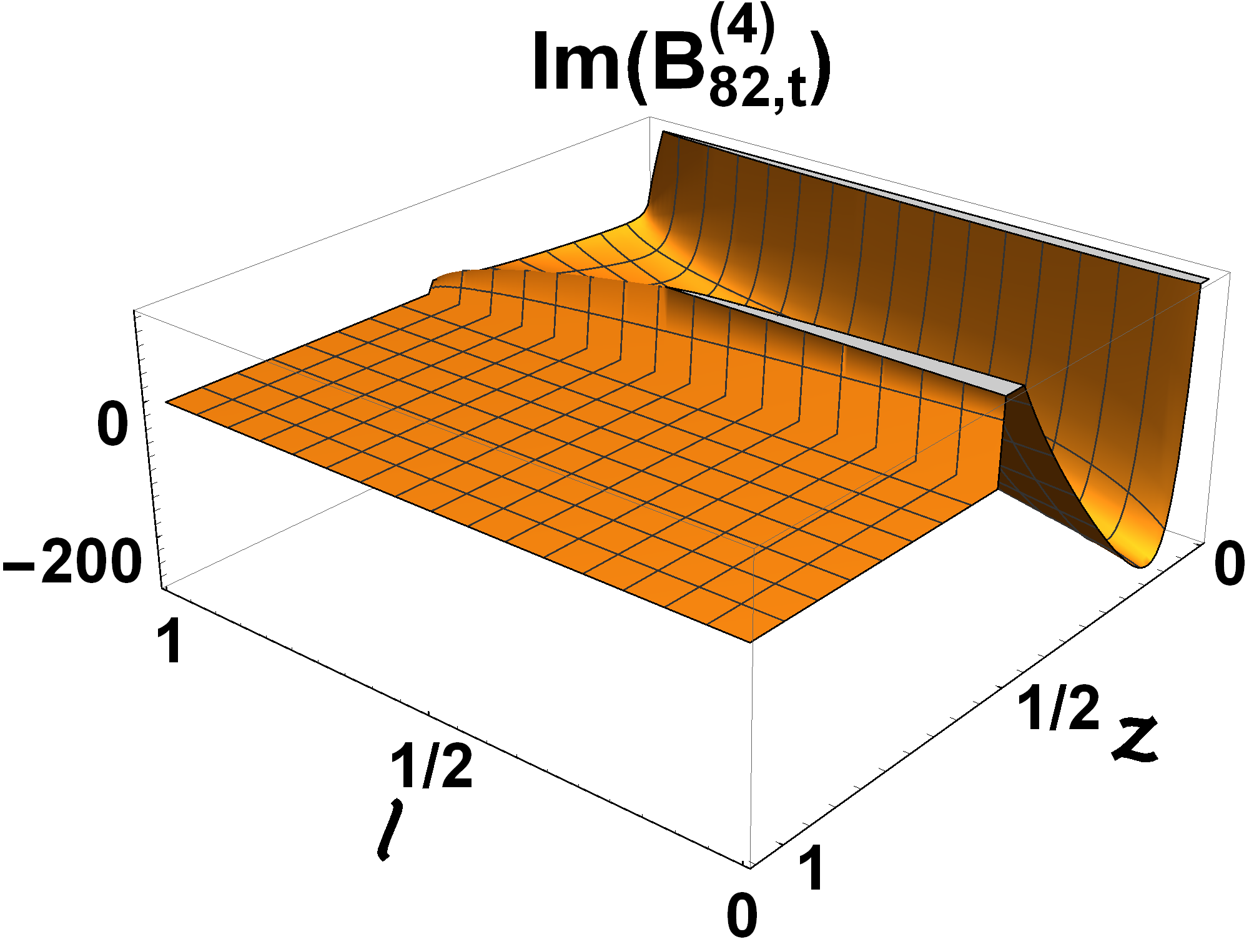} \\
        \imgBotRe{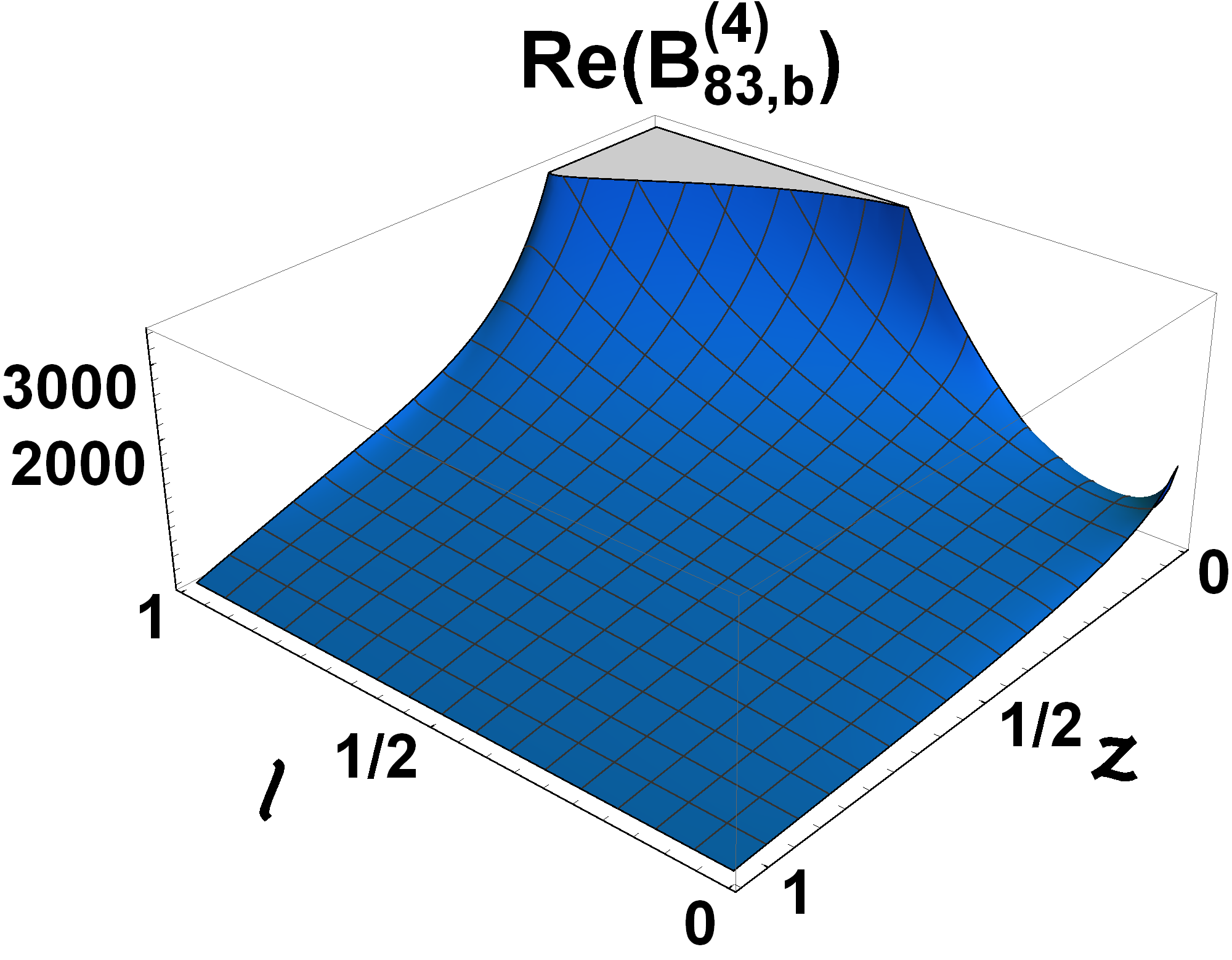} & \imgBotIm{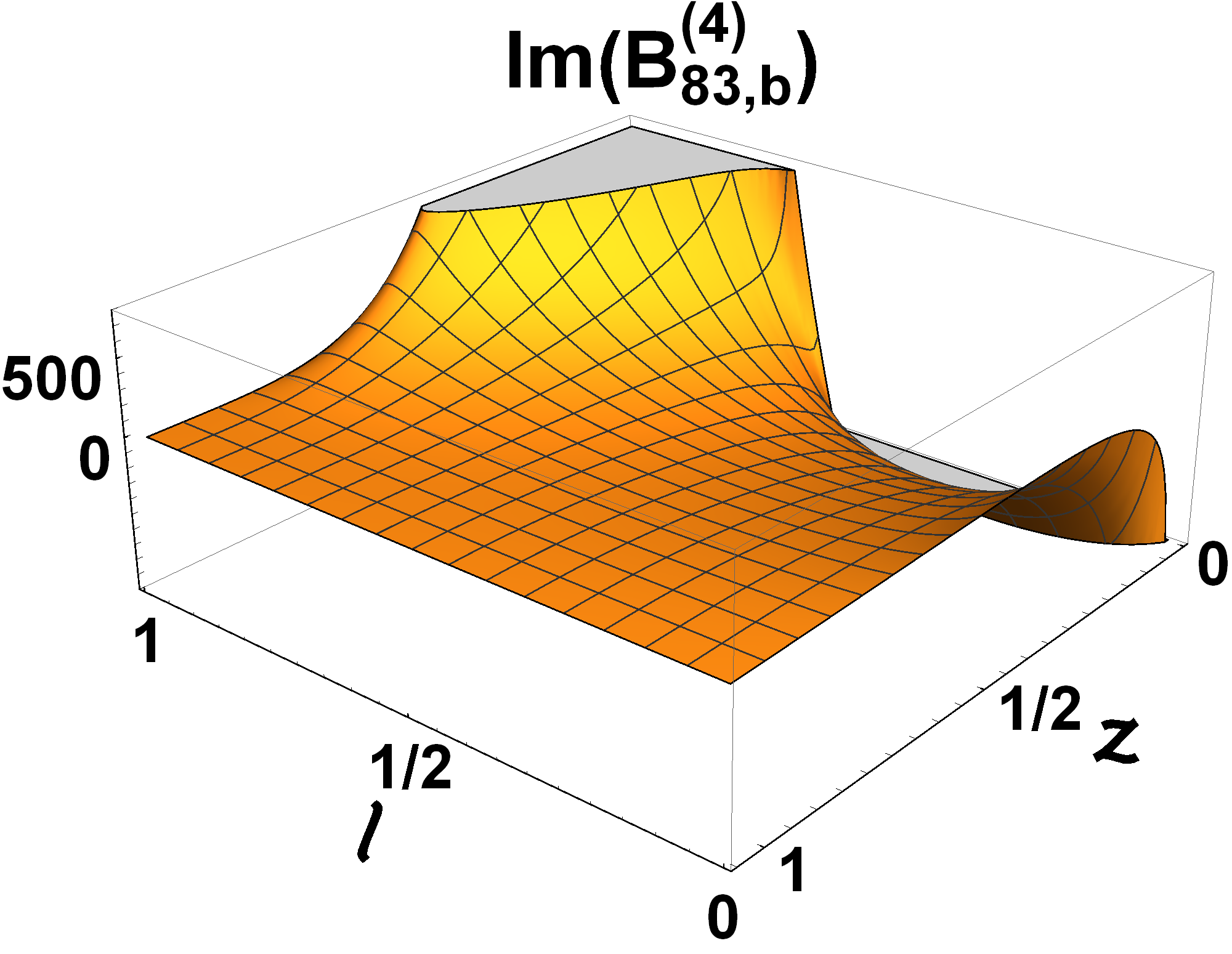} & \imgTopRe{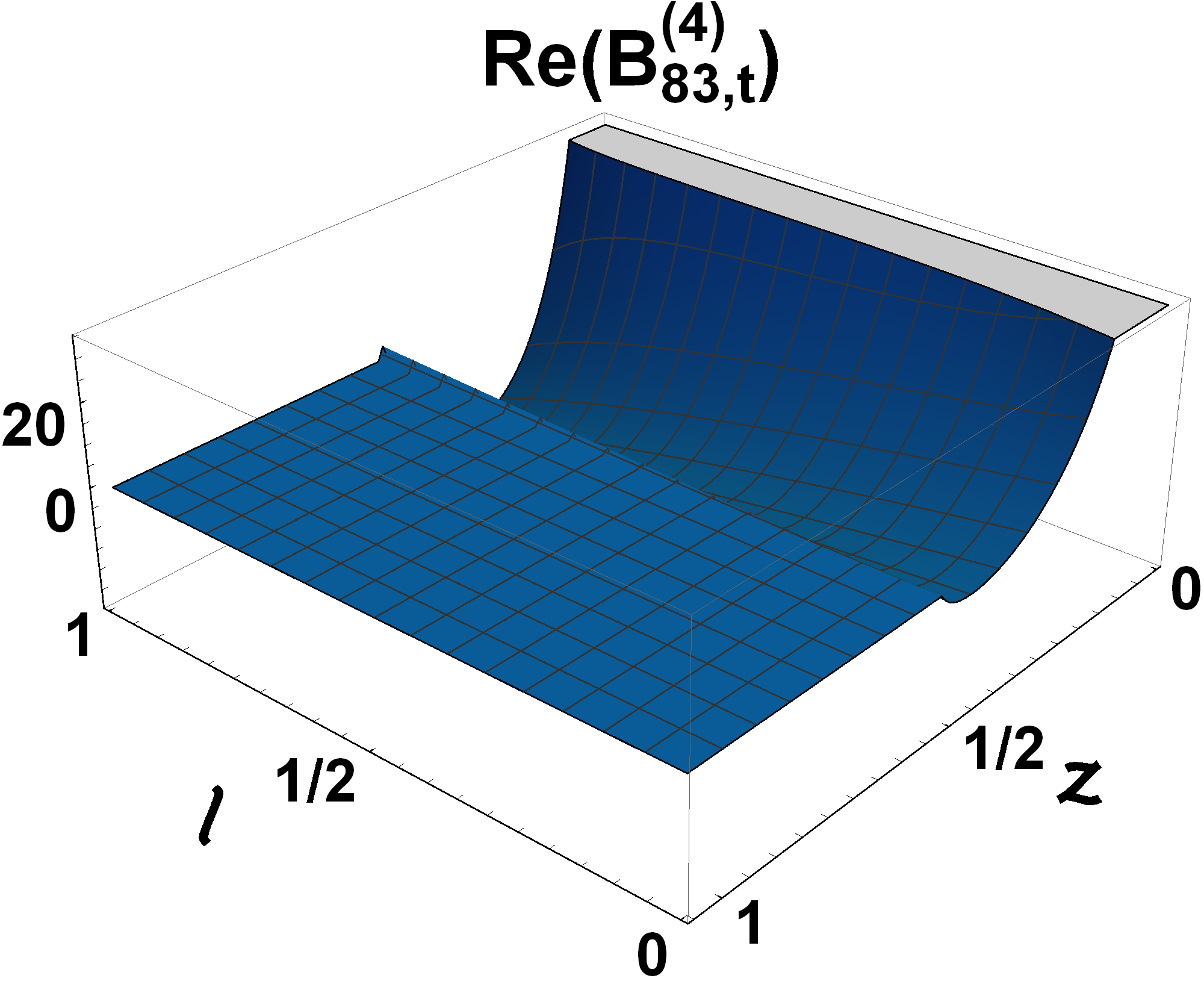} & \imgTopIm{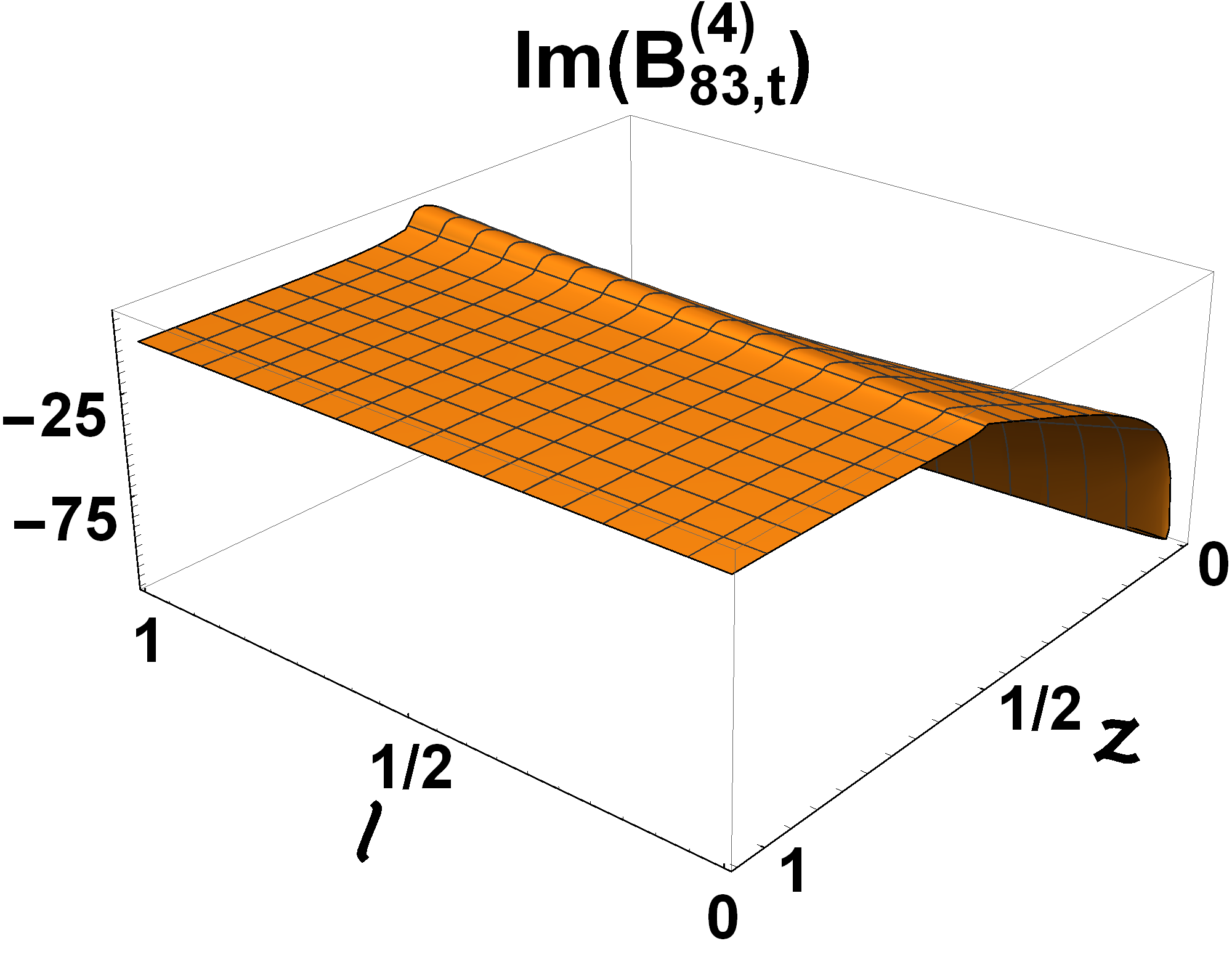} \\
        \imgBotRe{83} & \imgBotIm{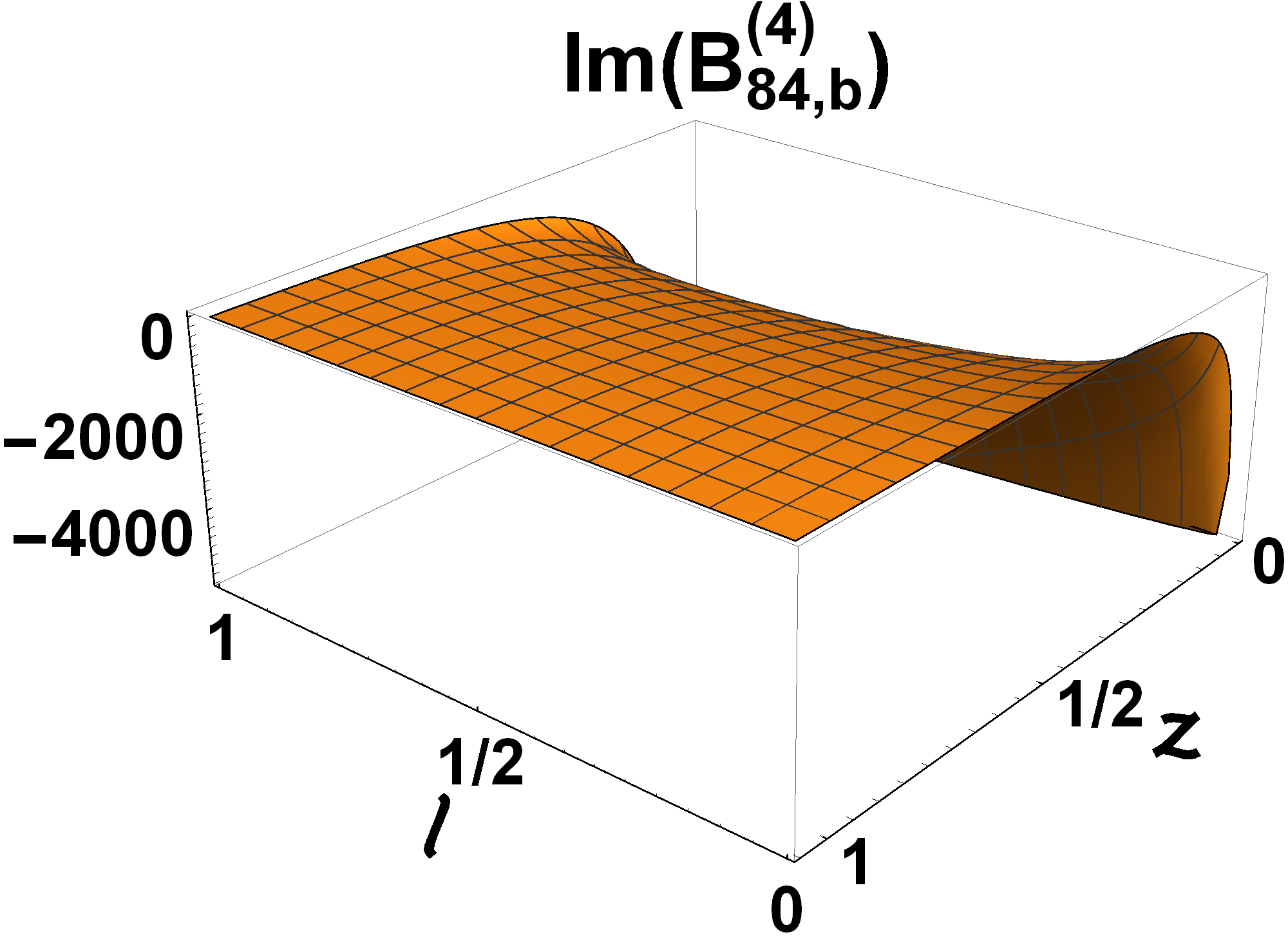} & \imgTopRe{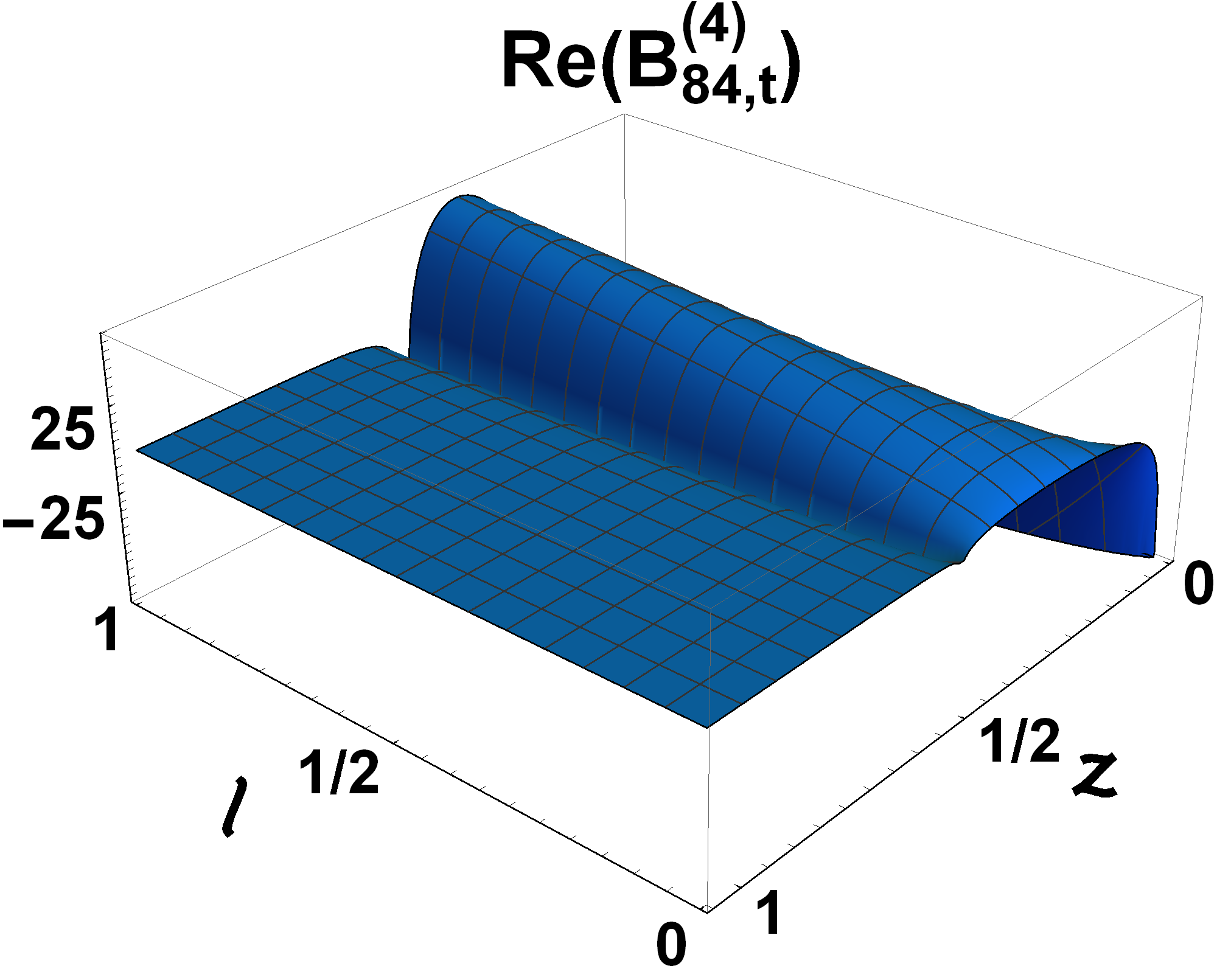} & \imgTopIm{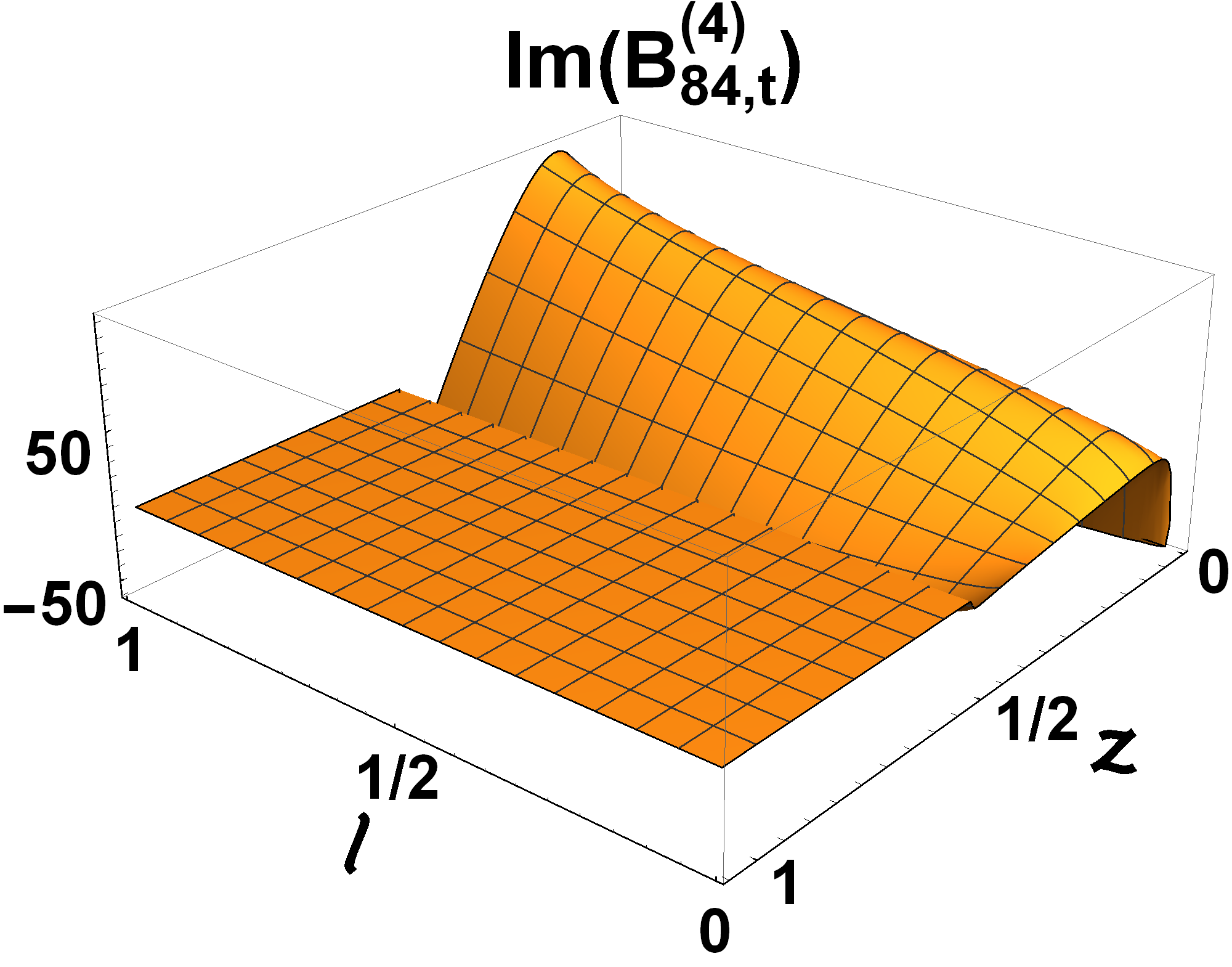} \\
 \end{tabular}
 \caption{On the left (resp. right) are shown the real (blue) and imaginary (orange) part of the integrals of the top sector of Family G in the case of a virtual bottom (resp. top) quark running in the loop.}
 \label{fig:plots}
\end{figure}
%\FloatBarrier
In this section we present explicit results that were obtained using the expansion method described in the previous section. Specifically, we used our method to compute the integrals in $10000$ points covering the physical region given below, for both the top- and bottom-quark corrections, and we present plots thereof. 
%The relevant physical region for Higgs plus jet production is given by,
We compute the integrals in the physical region given by
\begin{align}
    \label{eq:physicalregionst}
    s > 0\,,\quad t<0 \,,\quad s + t - p_4^2 > 0\,.
\end{align}
We may map that region to the unit square by using the parametrization,
\begin{align}
    s = \frac{p_4^2}{z}\,,\quad t = \frac{p_4^2 \, l \, (z-1)}{z}\,.
\end{align}
Since we chose to work with $m^2=1$, the value for $p_4^2$ is given by $m_H^2/m_q^2$  where $m_H$ denotes the mass of the Higgs particle, and $m_q$ denotes the mass of the internal quark. For the top quark we approximate the ratio by $p_4^2 = 13/25$, while for the bottom quark we consider the ratio $p_4^2 = 323761/361$. 

For the case of the top quark, the particle production threshold $s=4m^2$ corresponds to  $z=13/100$. For the sake of the presentation of the plots, we use a M{\"o}bius transformation to map $z=13/100$ to $1/2$, while keeping $z = 0$ and $z=1$ fixed. Thus, we consider the following parametrizations of the physical regions of the top and bottom quark contributions,
\begin{align}
    \text{top}~~~(l,z)_t:\quad s &= \frac{87-74 z}{25 z}\,,\,&&t = \frac{87 \, l \, (z-1)}{25 z}\,,\, &&p_4^2 = \frac{13}{25}\,, \nonumber\\
    \text{bottom}~~~(l,z)_b:\quad s &= \frac{323761}{361 z}\,,\,&&t = \frac{323761 \, l \, (z-1)}{361 z}\,,\, &&p_4^2 = \frac{323761}{361}\,.
\end{align}
To produce plots in these regions we seek to compute $n^2$ evenly spaced points on the unit square for all basis integrals, and in particular we let $n = 100$, so that we obtain 10000 points in total. We explain next how we obtained results in these points. For convenience we use the notation $a\rightarrow b$ to denote a line, we denote coordinates in the physical regions by pairs $(l,z)$, and we denote the heavy mass limit by $\vec{0}$. The following discussion applies to both the top and bottom region, given their respective set of $(l,z)$-coordinates. 

All of the results are derived using series expansions up to order $\mathcal{O}(\lambda^{50})$. First we set $k=3$, and move from the heavy mass limit to the point $(1/(n+1),n/(n+1))$. Then, we continue by moving along a vertical line $(1/(n+1),n/(n+1)) \rightarrow (1/(n+1), 1/(n+1))$. This vertical line may be used to obtain values at the points $(1/(n+1), y/(n+1))$ for $y=1,\ldots,n$. We may then consider $n$ horizontal lines $(1/(n+1),x/(n+1)) \rightarrow (n/(n+1), x/(n+1))$ for $x=1,\ldots,n$, to obtain values at the points $(x/(n+1),y/(n+1))$, for $x,y = 1,\ldots,n$. The situation is depicted in Fig. \ref{fig:topregion}, for the simpler case where $n=10$. We computed the expansions along the horizontal lines with $k=2$, in order to reduce the number of line segments needed and to save computation time. By working with $k=3$ for the first two lines, we made sure that the precision of the expansions along the horizontal lines is not limited by the precision of the expansions along the first two lines.

The resulting plots of integrals $B_{80}-B_{84}$ for the top- and the bottom-quark are provided in Fig. \ref{fig:plots}. Note that as $l$ and $z$ range from zero to one, we travel across the full physical region defined in Eq. (\ref{eq:physicalregionst}). For the plots we let $n=100$, and therefore the variables $l$ and $z$ range from $1/101$ to $100/101$. Thus, in the plots a small part of the physical region is cut off at the boundary. In terms of the variables $s$ and $t$, the plotted regions are given by:
\begin{align}
    \text{top}:~~~ &\left(\frac{1387}{2500} \leq s \leq \frac{8713}{25} \right) \cup \left(\frac{52}{101}-\frac{100 }{101}s \leq t \leq \frac{13}{2525}-\frac{s}{101}\right)\,,\nonumber\\
    \text{bottom}:~~~ & \left(\frac{32699861}{36100} \leq s \leq \frac{32699861}{361}\right) \cup \left(\frac{32376100}{36461}-\frac{100}{101}s \leq t \leq \frac{323761}{36461}-\frac{s}{101}\right)\,.
\end{align}
Note that it is also possible to obtain numerical samples at points on the boundary of the physical region where the integrals are finite, see for example Ref. \cite{Abreu:2020jxa}.

We computed the boundary data for the horizontal lines of the top and bottom physical regions on a laptop using a single core. We computed all horizontal lines on a cluster with 48 cores. The run for the horizontal lines of the top quark and the run for the horizontal lines of the bottom quark, both took a few hours to complete on the cluster.

\subsection{Cross-checks of the expansions}

We have performed several checks of our results. The first class of cross-checks was performed by evaluating multiple points by reaching them along different contours. The error that is accumulated while transporting results is different based on the chosen contour. Therefore, the difference of the results obtained through different contours gives a very good estimate of the precision of the results. In Table \ref{tab:internalcrosschecks} we present the results of a number of cross-checks that were performed in this way. The maximum relative error that we encountered among all the points that were checked, is of order $\mathcal{O}(10^{-25})$, indicating that the results are valid up to at least 25 significant digits.
\begin{table}[h!]
\centering
\begin{tabular}{c c c c} 
 \hline
 \text{Line(s)}. & Evaluated at & \text{\#Segments} $(k=2)$ & \text{Max relative error} \\ [0.5ex] 
 \hline\hline
 $\vec{0} \rightarrow \left(\frac{1}{101}, \frac{1}{101}\right)_t$ & Endpoint & 16 & $\mathcal{O}(10^{-28})$ \\
 $\vec{0} \rightarrow \left(\frac{1}{101}, \frac{1}{101}\right)_b$ & Endpoint & 31 & $\mathcal{O}(10^{-26})$ \\
 $\vec{0} \rightarrow\, \icol{s = 53\\t = -11\\p_4^2 = 23} \rightarrow \left(\frac{100}{101}, \frac{45}{101}\right)_b$ & Endpoint& 47 & $\mathcal{O}(10^{-25})$ \\
 $\underset{\text{for }\,x=1,\ldots,100}{\left(\frac{x}{101}, \frac{100}{101}\right)_t \rightarrow \left(\frac{x}{101}, \frac{1}{101}\right)_t}$ & $\underset{\text{for }\, x,y=1,\ldots,100}{\left(\frac{x}{101}, \frac{y}{101}\right)_t}$ & 2568 & $\mathcal{O}(10^{-25})$ \\
 $\left(\frac{45}{101}, \frac{45}{101}\right)_b \rightarrow \left(\frac{1}{101}, \frac{100}{101}\right)_t$ & Endpoint & 21 & $\mathcal{O}(10^{-27})$ \\\hline
\end{tabular}
\caption{This table presents a number of internal cross-checks of our results. In the first column we give additional lines along which we computed results, different from the lines in Fig. \ref{fig:topregion}. These results were then compared to the results that we generated for the plots, which were computed in the manner illustrated in Fig. \ref{fig:topregion}. For the lines starting from $\vec{0}$, we fixed the boundary conditions in the heavy mass limit, while for the lines in the last two rows we fixed the boundary conditions from the results that we generated for the plots. In the last column we give highest value of the relative error $\left|B_{i,\text{cross-check}}^{(\epsilon\text{-order})}/B_{i,\text{plot}}^{(\epsilon\text{-order})}\right| - 1$ for all integrals $i = 1,\ldots,84$, and $\epsilon$-orders 0 to 4.}
\label{tab:internalcrosschecks}
\end{table}

For the top-quark integrals we compared our results against FIESTA \cite{Smirnov:2015mct} for multiple points of the physical region finding full agreement within the Monte Carlo error reported by FIESTA. For the physical region of the bottom quark we checked most of the integrals against FIESTA and SecDec \cite{Borowka:2017idc}. However, for some of the integrals, these programs encounter numerical instabilities. In those cases we have performed different checks.
Firstly we cross checked our results against FIESTA in the point $(s = 53,\, t = -11,\, p_4^2 = 23, m^2 = 1)$ finding full agreement. This provides a direct check of the analytic continuation past the thresholds $s=4m^2$ and $p_4^2 = 4m^2$.
In addition, we have performed a numerical cross-check against a private code \cite{Capatti:2019edf} for the numerical evaluation of multi-loop integrals in momentum space using the loop tree duality \cite{Capatti:2019ypt} (for related work on the loop tree duality see also \cite{Catani:2008xa,Aguilera-Verdugo:2019kbz,Bierenbaum:2010cy,Runkel:2019yrs}). In particular, we compared integrals $B_{72}$ and $B_{76}$ in the point $(10/101,10/101)_b = (s = 32699861/3610, t = -29462251/36461, p_4^2 = 323761/361, m^2 = 1)$ finding full agreement.
Lastly, for one of the internal cross-checks we transported the results for the bottom-quark integrals from the point $\left(\frac{45}{101},\frac{45}{101}\right)_b$ along a straight line to the point $\left(\frac{1}{101},\frac{100}{101}\right)_t$. We compared this to the results that were obtained in the point $\left(\frac{1}{101},\frac{100}{101}\right)_t$ by transporting directly from the heavy mass limit, and found a relative deviation of $\mathcal{O}(10^{-27})$. This cross-check is indicated by the last row in Table \ref{tab:internalcrosschecks}.

\section{Conclusion}
\label{sec:concl}
In this paper we computed a family of two-loop non-planar master integrals relevant for the QCD corrections to Higgs plus one jet production in the full theory. Our result, together with \cite{Bonciani:2016qxi,Bonciani:2019jyb,Francesco:2019yqt}, provide the full set of master integrals required for the computation of the NLO corrections to Higgs plus one jet production, and the NLO corrections to the $p_T$ distribution of the Higgs. Moreover, our results provide the full set of master integrals relevant for the NLO corrections to Higgs decay to three partons, and the single-real radiation contributions to the NNLO corrections to inclusive Higgs production.

The computation was performed by using the differential equations method. More specifically, we defined an integral basis such that most of the integrals satisfy differential equations in canonical form. Three integral sectors are coupled, and their solution involve functions of elliptic type. Having phenomenological applications in mind, we solved the differential equations along contours in the space of kinematic invariants, in terms of one-dimensional generalized power series. More specifically, given a boundary point where the value of the integrals is known, we defined the differential equations along a contour connecting a boundary point to a new point of the kinematic regions. In this way the problem was effectively reduced to one with a single scale, and finding the series solution was algorithmic. We showed that this method is efficient, and can be repeated in order to compute the integrals in any point of the kinematic regions. The analytic continuation of the series solution across the physical thresholds is straightforward, as it requires only the analytic continuation of logarithms and square roots.

In order to show the generality of our approach, we computed the master integrals for both the top- and bottom-quark mass. Moreover, we explicitly obtained results for a large set of points covering our physical regions. The typical evaluation time is of the order of $1$ second per integral, with a relative accuracy of order $10^{-24}$, on a single CPU core. If needed, the numerical precision can be made arbitrarily high by increasing the truncation order of the power series. These features render our methods well suited for Monte Carlo phase-space integrations.

We remark that the applicability of our methods does not rely on the number of physical scales, specific kinematic configurations, or a particular form of the differential equations, and a public implementation has been recently developed in \cite{Hidding:2020ytt}. For this reasons, we believe that our approach will be relevant for the computation of several processes of phenomenological interest.

\section*{Acknowledgements}

The authors would like to thank Zeno Capatti and Valentin Hirschi for performing cross checks of our results against a private code for the numerical evaluation of multi-loop integrals using the loop tree duality. We thank Roberto Bonciani, Vittorio Del Duca, Johannes Henn and Volodya Smirnov for useful discussions and collaboration on related work.
The work of HF is part of the HiProLoop project funded by the European Union's Horizon 2020 research and innovation programme under the Marie Sk{\l}odowska-Curie grant agreement 747178.
The work of FM was funded by the European Research Council (ERC) under grant agreement No. 694712 (pertQCD) and by the Swiss National Science Foundation project No. 177632 (ElliptHiggs). The work of MH was funded by the European Research Council (ERC) under grant agreement No. 647356 (CutLoops). The work of LM received funding from the European Research Council (ERC) under the European Union's Horizon 2020 research and innovation programme, {\it Novel structures in scattering amplitudes} (grant agreement No. 725110).
GS is supported by the Simons Investigator Award ${\#}$376208 of A. Volovich. MH, LM and GS would like to thank ETH Zurich for hospitality during the preparation of this work.

\vfill

\appendix

\section{Canonical basis and basis for elliptic sectors}
In this section we provide the set of 84 basis integrals used in this paper, written in terms of the set of master integrals depicted in Figure \ref{fig:masters} and defined as in Eq. (\ref{Is}). 

The canonical basis for the first 71 integrals is,
\begin{align*}
B_{1}&=\epsilon ^2 \text{I}_{0,2,0,0,2,0,0,0,0} \,,\\
B_{2}&=\epsilon ^2 r_2 r_6  \text{I}_{0,2,0,0,2,0,1,0,0} \,,\\
B_{3}&=\epsilon ^2r_1 r_5  \text{I}_{0,2,0,1,0,2,0,0,0} \,,\\
B_{4}&=\epsilon ^2 s  \text{I}_{1,2,0,0,2,0,0,0,0} \,,\\
B_{5}&=\epsilon ^2 r_2 r_6 \left(\frac{1}{2} \text{I}_{1,2,0,0,2,0,0,0,0}+\text{I}_{2,2,0,0,1,0,0,0,0} \right)\,,\\
B_{6}&=\epsilon ^2 t  \text{I}_{0,2,1,0,0,2,0,0,0} \,,\\
B_{7}&=\epsilon ^2 r_3 r_7 \left(\frac{1}{2}   \text{I}_{0,2,1,0,0,2,0,0,0}+ \text{I}_{0,2,2,0,0,1,0,0,0}\right) \,,\\
B_{8}&=\epsilon ^2  (p_4^2-s-t)\text{I}_{1,0,0,2,0,2,0,0,0} \,,\\
B_{9}&=\epsilon ^2r_4 r_8\left(\frac{1}{2}  \text{I}_{1,0,0,2,0,2,0,0,0}+ \text{I}_{2,0,0,1,0,2,0,0,0}\right) \,,\\
B_{10}&= \epsilon ^2 p_4^2 \text{I}_{0,2,1,0,2,0,0,0,0} \,,\\
B_{11}&= \epsilon ^2 r_1 r_5 \left(\frac{1}{2}  \text{I}_{0,2,1,0,2,0,0,0,0}+ \text{I}_{0,2,2,0,1,0,0,0,0} \right)\,,\\
B_{12}&=\epsilon ^3 s \text{I}_{0,2,0,0,1,1,1,0,0} \,,\\
B_{13}&=-\epsilon ^2r_1 r_2 r_5 r_6 \text{I}_{0,2,0,1,2,0,1,0,0} \,,\\
B_{14}&=\epsilon ^3 s\text{I}_{1,2,0,0,1,1,0,0,0} \,,\\
B_{15}&=\epsilon ^3 t\text{I}_{0,2,1,0,0,1,1,0,0} \,,\\
B_{16}&=\epsilon ^3 (p_4^2-s-t) \text{I}_{1,0,0,2,1,1,0,0,0}\,,\\
B_{17}&=\epsilon ^3 (p_4^2-t) \text{I}_{0,2,1,0,1,1,0,0,0}\,,\\
B_{18}&=\epsilon ^3 (s+t)\text{I}_{1,0,0,2,0,1,1,0,0} \,,\\
B_{19}&=\epsilon ^3 (s-p_4^2)\text{I}_{0,2,1,0,1,0,1,0,0} \,,\\
B_{20}&=\epsilon ^2 m^2 (s-p_4^2) \text{I}_{0,3,1,0,1,0,1,0,0}  \,,\\
B_{21}&=\epsilon ^2\frac{r_2 r_6}{4 (s-2 p_4^2)} \left(4 \left(m^2 s+p_4^4-p_4^2 s\right)\text{I}_{0,2,1,0,2,0,1,0,0}+\right.\\
&+\left. 4 m^2 (s-p_4^2) \text{I}_{0,3,1,0,1,0,1,0,0} +6 \epsilon  (p_4^2-s)\text{I}_{0,2,1,0,1,0,1,0,0} -3 p_4^2 \text{I}_{0,2,1,0,2,0,0,0,0}\right)\,,\\
B_{22}&=\epsilon ^3 (p_4^2-s) \text{I}_{1,1,0,1,2,0,0,0,0}  \,,\\
B_{23}&=\epsilon ^2 m^2 (p_4^2-s)\text{I}_{1,1,0,1,3,0,0,0,0} \,,\\
B_{24}&=\epsilon ^2 \frac{r_1 r_5}{4 (p_4^2-2 s)} \left(4 m^2 \text{I}_{1,1,0,1,3,0,0,0,0} (p_4^2-s)+4 m^2 p_4^2 \text{I}_{1,2,0,1,2,0,0,0,0}\right.+\\
&\left.+6 \epsilon  (s-p_4^2)\text{I}_{1,1,0,1,2,0,0,0,0} -4 p_4^2 s \text{I}_{1,2,0,1,2,0,0,0,0}+\right.\\
&+\left.4 s^2 \text{I}_{1,2,0,1,2,0,0,0,0}-3 s \text{I}_{1,2,0,0,2,0,0,0,0}\right)\,,\\
B_{25}&=\epsilon^3 (p_4^2-t)\text{I}_{0,1,1,1,0,2,0,0,0} \,,\\
B_{26}&=\epsilon^2 m^2  (p_4^2-t) \text{I}_{0,1,1,1,0,3,0,0,0}  \,,\\
B_{27}&=\epsilon^2 \frac{r_1 r_5  }{4 (p_4^2-2 t)}\left(4 m^2  (p_4^2-t)\text{I}_{0,1,1,1,0,3,0,0,0} +4 m^2 p_4^2 \text{I}_{0,2,1,1,0,2,0,0,0}\right.\\
&+\left.6 \epsilon  (t-p_4^2) \text{I}_{0,1,1,1,0,2,0,0,0} -4 p_4^2 t \text{I}_{0,2,1,1,0,2,0,0,0}+\right.\\
&+\left.4 t^2 \text{I}_{0,2,1,1,0,2,0,0,0}-3 t \text{I}_{0,2,1,0,0,2,0,0,0}\right) \,,\\
B_{28}&=\epsilon ^3 (s+t) \text{I}_{1,1,0,1,0,2,0,0,0}  \,,\\
B_{29}&=\epsilon^2 m^2 (s+t)\text{I}_{1,1,0,1,0,3,0,0,0} \,,\\
B_{30}&=-\epsilon^2\frac{r_1 r_5  }{4 (p_4^2-2 (s+t))}\left(4 (m^2 p_4^2-(s+t) (p_4^2-s-t))\text{I}_{1,1,0,2,0,2,0,0,0}\right.\\
&+\left. 4 m^2  (s+t)\text{I}_{1,1,0,1,0,3,0,0,0}+3  (-p_4^2+s+t)\text{I}_{1,0,0,2,0,2,0,0,0}-\right.\\
&-\left. 6 \epsilon   (s+t)\text{I}_{1,1,0,1,0,2,0,0,0} \right) \,,\\
B_{31}&= \epsilon ^3 s r_1 r_5  \text{I}_{0,2,0,1,1,1,1,0,0} \,,\\
B_{32}&=\epsilon ^4 (p_4^2-t)\text{I}_{0,1,1,1,1,1,0,0,0}  \,,\\
B_{33}&=\epsilon ^3 (p_4^2-t)r_1 r_5  \text{I}_{0,2,1,1,1,1,0,0,0}\,,\\
B_{34}&=\epsilon ^4 (s+t)\text{I}_{1,1,0,1,0,1,1,0,0}\,,\\
B_{35}&=\epsilon ^3  (s+t)r_1 r_5 \text{I}_{1,1,0,2,0,1,1,0,0} \,,\\
B_{36}&=\epsilon ^4  (p_4^2-s-t)\text{I}_{1,1,1,0,1,1,0,0,0} \,,\\
B_{37}&=-\epsilon^3 r_2 r_3 r_9  \text{I}_{1,2,1,0,1,1,0,0,0}\,,\\
B_{38}&=\epsilon^4 t \text{I}_{1,0,1,1,0,1,1,0,0} \,,\\
B_{39}&=-\epsilon^3 r_2 r_4 r_{10} \text{I}_{1,0,1,2,0,1,1,0,0} \,,\\
B_{40}&=\epsilon^4(p_4^2-s) \text{I}_{1,1,1,1,0,1,0,0,0}  \,,\\
B_{41}&= \epsilon^3  r_3 r_4 r_{11}\text{I}_{1,1,1,1,0,2,0,0,0}\,,\\
B_{42}&=\frac{1}{4} \epsilon ^2 \left(4 \epsilon \frac{1}{t}\left(m^2 (p_4^2-s)^2+p_4^2 t (-p_4^2+s+t)\right) \text{I}_{1,1,1,2,0,1,0,0,0} \right.+\\
&+\left. 2 \epsilon (2 m^2 (p_4^2-s)+t (-p_4^2+s+t))\text{I}_{1,1,1,1,0,2,0,0,0}+\right.\\
&+\left.\frac{3 (-2 m^2 p_4^2+2 m^2 s+p_4^2 t)}{p_4^2-2 t}\text{I}_{0,2,1,0,0,2,0,0,0}\right.-\\
&-6 \epsilon\left.\frac{(p_4^2-t) (2 m^2 (p_4^2-s)-p_4^2 t)}{t (p_4^2-2 t)}\text{I}_{0,1,1,1,0,2,0,0,0}\right.+\\
&+\left.\frac{4 m^2 (p_4^2-t) (2 m^2 (p_4^2-s)-p_4^2 t)}{t (p_4^2-2 t)}\text{I}_{0,1,1,1,0,3,0,0,0}\right.+\\
&+\left.\frac{4 (m^2 p_4^2+t (t-p_4^2)) (2 m^2 (p_4^2-s)-p_4^2 t)}{t (p_4^2-2 t)}\text{I}_{0,2,1,1,0,2,0,0,0}\right.+\\
&+6 \epsilon\left.\frac{  (s+t) (-2 m^2 p_4^2+2 m^2 s+p_4^2 t)}{t (p_4^2-2 (s+t))} \text{I}_{1,1,0,1,0,2,0,0,0}\right.+\\
&+\left.\frac{3  (p_4^2-s-t) (-2 m^2 p_4^2+2 m^2 s+p_4^2 t)}{t (p_4^2-2 (s+t))}\text{I}_{1,0,0,2,0,2,0,0,0}\right.-\\
&-\left.\frac{4 m^2 (s+t) (-2 m^2 p_4^2+2 m^2 s+p_4^2 t)}{t (p_4^2-2 (s+t))}\text{I}_{1,1,0,1,0,3,0,0,0}\right.+\\
&+\left.\frac{4 (2 m^2 (p_4^2-s)-p_4^2 t) (m^2 p_4^2-(s+t) (p_4^2-s-t))}{t (p_4^2-2 (s+t))}\text{I}_{1,1,0,2,0,2,0,0,0} \right) \,,\\
B_{43}&=\epsilon^4 (p_4^2-s) \text{I}_{0,1,1,1,1,0,1,0,0}  \,,\\
B_{44}&=\epsilon^3 (s-p_4^2) r_2 r_6\text{I}_{0,1,1,1,1,0,2,0,0}  \,,\\
B_{45}&=\epsilon^3 (p_4^2-s)r_1 r_5 \text{I}_{0,2,1,1,1,0,1,0,0}  \,,\\
B_{46}&=\epsilon^2\left(m^2 (p_4^2-s)^2 \text{I}_{0,2,1,1,1,0,2,0,0}-2 (2 m^2 p_4^2+2 m^2 s-p_4^2 s)\text{I}_{0,1,0,2,1,0,2,0,0} \right.+\\
&+ \left. \epsilon (p_4^2-s)\left(s\text{I}_{0,1,1,1,1,0,2,0,0}- p_4^2\text{I}_{0,2,1,1,1,0,1,0,0}\right)\right) \,,\\
B_{47}&=-\epsilon^3 r_2 r_{14} \text{I}_{0,2,1,0,1,1,1,0,0}\,,\\
B_{48}&=-\epsilon ^2 r_2 r_3 r_9 \left(m^2 \text{I}_{0,3,1,0,1,1,1,0,0}-\epsilon  \text{I}_{0,2,1,0,1,1,1,0,0}\right)\,,\\
B_{49}&=s \epsilon ^3 \left((-m^2-p_4^2+s)\text{I}_{0,2,1,0,1,1,1,0,0}-\text{I}_{-1,2,1,0,1,1,1,0,0}+\text{I}_{0,2,0,0,1,1,1,0,0}-\right.\\
&-\left. \text{I}_{0,2,1,-1,1,1,1,0,0}+\text{I}_{0,2,1,0,1,1,0,0,0}+\text{I}_{0,2,1,0,1,1,1,-1,0}\right) \,,\\
B_{50}&=-\epsilon ^3 r_2 r_{15}  \text{I}_{1,0,0,2,1,1,1,0,0} \,,\\
B_{51}&=-\epsilon ^2 r_2 r_4 r_{10}  \left(m^2 \text{I}_{1,0,0,3,1,1,1,0,0}-\epsilon  \text{I}_{1,0,0,2,1,1,1,0,0}\right) \,,\\
B_{52}&=\epsilon ^3 s \left(-m^2 \text{I}_{1,0,0,2,1,1,1,0,0}+\text{I}_{0,0,0,2,1,1,1,0,0}-\text{I}_{1,0,-1,2,1,1,1,0,0}+\text{I}_{1,0,0,1,1,1,1,0,0}+\right.\\
&+\left.\text{I}_{1,0,0,2,0,1,1,0,0}-\text{I}_{1,0,0,2,1,1,1,-1,0}\right) \,,\\
B_{53}&=\epsilon ^4 \text{I}_{0,1,1,1,0,1,1,0,0} (s+t) \,,\\
B_{54}&=\epsilon ^3 r_1 r_5  \left(s \text{I}_{0,1,1,2,0,1,1,0,0}-t \text{I}_{0,2,1,1,0,1,1,0,0}\right) \,,\\
B_{55}&=-\epsilon ^3r_2 r_3 r_9  \text{I}_{0,1,2,1,0,1,1,0,0}\,,\\
B_{56}&=\frac{1}{2} \epsilon ^3 \left( (2 m^2 s+2 m^2 t-p_4^2 s)\text{I}_{0,1,1,2,0,1,1,0,0}+(2 m^2 s+2 m^2 t-p_4^2 t)\text{I}_{0,2,1,1,0,1,1,0,0} \right) \,,\\
B_{57}&=\epsilon^2\left(2 (p_4^2+s)\left(m^2\text{I}_{0,1,1,1,0,3,0,0,0} - \epsilon \text{I}_{0,1,1,1,0,2,0,0,0}\right)\right.+\\
&+\left.\epsilon  (s+t)\left((p_4^2-s)\text{I}_{0,1,2,1,0,1,1,0,0}+ \text{I}_{0,0,2,1,0,1,1,0,0}+\right.+\left.\text{I}_{0,1,2,0,0,1,1,0,0}-\text{I}_{0,1,2,1,0,1,1,-1,0}\right.\right) \,,\\
B_{58}&=\epsilon ^4(p_4^2-t) \text{I}_{1,1,0,1,1,1,0,0,0} \,,\\
B_{59}&=\epsilon ^3 r_1 r_5 \left((-p_4^2+s+t)\text{I}_{1,1,0,2,1,1,0,0,0}+s \text{I}_{1,2,0,1,1,1,0,0,0}\right) \,,\\
B_{60}&=-\epsilon ^3r_2 r_4 r_{10}  \text{I}_{2,1,0,1,1,1,0,0,0} \,,\\
B_{61}&=\frac{1}{2} \epsilon ^3 \left((2 m^2 (p_4^2 - t) + p_4^2 (-p_4^2 + s + t))\text{I}_{1,1,0,2,1,1,0,0,0}+\right.\\
&+\left.(2 m^2 p_4^2-2 m^2 t-p_4^2 s)\text{I}_{1,2,0,1,1,1,0,0,0} \right) \,,\\
B_{62}&=-\epsilon ^2 \left(-2 m^2 (p_4^2+s) \text{I}_{1,1,0,1,0,3,0,0,0}+\right.\\
&+\left.2 p_4^2 \epsilon  \text{I}_{1,1,0,1,0,2,0,0,0}-p_4^2 \epsilon  \text{I}_{2,1,0,1,1,1,0,-1,0}+2 s \epsilon  \text{I}_{1,1,0,1,0,2,0,0,0}+t \epsilon  \text{I}_{2,1,0,1,1,1,0,-1,0}\right) \,,\\
B_{63}&=\epsilon ^4 \text{I}_{1,1,1,1,1,0,1,0,0} (s-p_4^2)^2 \,,\\
B_{64}&=r_3 r_{12} \epsilon ^4 \text{I}_{1,1,1,1,0,1,1,0,0}\,,\\
B_{65}&=\epsilon^2\left\{\epsilon \frac{(3 m^2 (p_4^2-s)+t (-2 p_4^2+s+t))}{p_4^2-2 (s+t)} \text{I}_{1,1,0,1,0,2,0,0,0} \right.+\\
&+\left. \frac{ (p_4^2-s-t) (3 m^2 (p_4^2-s)+t (-2 p_4^2+s+t))}{2 (s+t) (p_4^2-2 (s+t))}\text{I}_{1,0,0,2,0,2,0,0,0}\right.-\\
&-\left. \frac{(2 m^2 (p_4^2-s)-p_4^2 t) (m^2 p_4^2-(s+t) (p_4^2-s-t))}{(s+t) (p_4^2-2 (s+t))} \text{I}_{1,1,0,2,0,2,0,0,0}\right.-\\
&-\left. \frac{1}{4 (p_4^2-2 t) (s+t)}\left[6 \epsilon  (p_4^2-t) (2 m^2 (s-p_4^2)+p_4^2 t)\text{I}_{0,1,1,1,0,2,0,0,0}+\right.\right.\\
&+\left.  4 m^2  (p_4^2-t) (2 m^2 (s-p_4^2)+p_4^2 t)\text{I}_{0,1,1,1,0,3,0,0,0}\right.+\\
&+\left. t (6 m^2 (p_4^2-s)+t (4 t-5 p_4^2))\text{I}_{0,2,1,0,0,2,0,0,0} \right.-\\
&-\left. 2 (m^2 p_4^2+t (t-p_4^2)) (2 m^2 (p_4^2-s)-p_4^2 t)\text{I}_{0,2,1,1,0,2,0,0,0} \right]-\\
&-\left. \epsilon ^2 t (\text{I}_{1,1,0,1,0,1,1,0,0}+\text{I}_{1,1,1,1,-1,1,1,0,0}+(s-p_4^2)\text{I}_{1,1,1,1,0,1,1,0,0}) +\right.\\
&+\left. \epsilon \frac{1}{2(s+t)} \left[2 \epsilon t  (p_4^2+s+2 t)\text{I}_{0,1,1,1,0,1,1,0,0}+  2\epsilon  t (p_4^2-2 s-t)\text{I}_{1,1,1,1,0,1,0,0,0} \right.\right.+\\
&+\left.  2\epsilon  t (s-p_4^2)\text{I}_{1,1,1,1,0,1,1,0,-1} + t  (p_4^2 s-2 m^2 (s+t))\text{I}_{0,1,1,2,0,1,1,0,0}\right.+\\
&+\left. t  (p_4^2 t-2 m^2 (s+t)) \text{I}_{0,2,1,1,0,1,1,0,0} + t  (s-p_4^2)\text{I}_{1,1,0,1,2,0,0,0,0} \right.-\\
&-\left. \frac{1}{2} p_4^2 t\text{I}_{0,2,1,0,2,0,0,0,0} + t^2 \text{I}_{0,2,1,0,0,1,1,0,0}+s t \text{I}_{1,2,0,0,1,1,0,0,0}\right.-\\
&-\left. t  (2 m^2 (p_4^2-s)+t (-p_4^2+s+t))\text{I}_{1,1,1,1,0,2,0,0,0}\right.-\\
&-\left. (m^2 (p_4^2-s)^2+p_4^2 t (-p_4^2+s+t) \text{I}_{1,1,1,2,0,1,0,0,0} \right]+\\
&+\left. m^2  \left(\frac{2 m^2 (s-p_4^2)+p_4^2 t}{p_4^2-2 (s+t)}+t\right)\text{I}_{1,1,0,1,0,3,0,0,0} \right\}\,,\\
B_{66}&=-2 \epsilon ^4 ( (s-p_4^2)\text{I}_{1,1,1,1,0,1,1,0,-1}+  (p_4^2+t)\text{I}_{0,1,1,1,0,1,1,0,0}) \,,\\
B_{67}&=r_1 r_5 \epsilon^2 \left[-\frac{2 m^2  (s+t)}{p_4^2-2 (s+t)}\text{I}_{1,1,0,1,0,3,0,0,0} \right.-\\
&-\left. \frac{2 (m^2 p_4^2-(s+t) (p_4^2-s-t))}{p_4^2-2 (s+t)} \text{I}_{1,1,0,2,0,2,0,0,0}\right.+\\
&+\left. \frac{2 m^2 (t-p_4^2)}{p_4^2-2 t}\text{I}_{0,1,1,1,0,3,0,0,0}+\frac{3  (p_4^2-s-t)}{2 (p_4^2-2 (s+t))}\text{I}_{1,0,0,2,0,2,0,0,0}\right.-\\
&-\left. \frac{2  (m^2 p_4^2+t (t-p_4^2))}{p_4^2-2 t}\text{I}_{0,2,1,1,0,2,0,0,0} +\frac{3 t  }{2 p_4^2-4 t}\text{I}_{0,2,1,0,0,2,0,0,0}\right.+\\
&+\left.  \epsilon \left( \frac{3    (s+t)}{p_4^2-2 (s+t)}\text{I}_{1,1,0,1,0,2,0,0,0}+ (s-p_4^2)(\text{I}_{1,1,1,2,0,1,0,0,0}+ \text{I}_{1,1,1,2,0,1,1,0,-1})\right.\right.+\\
&+\left.\left.2  t(\epsilon \text{I}_{1,1,1,1,0,1,1,0,0}+  \text{I}_{0,1,2,1,0,1,1,0,0}- \text{I}_{0,2,1,1,0,1,1,0,0}+ \text{I}_{1,0,1,2,0,1,1,0,0})\right.\right.+\\
&+\left.\left. p_4^2 \text{I}_{0,1,1,2,0,1,1,0,0}+  \frac{3 (p_4^2-t)}{p_4^2-2 t}\text{I}_{0,1,1,1,0,2,0,0,0}\right)\right] \,,\\
B_{68}&=\epsilon ^4 \left((p_4^2-s-t)\text{I}_{1,1,1,1,1,1,-1,0,0} - (p_4^2-s) (p_4^2-s-t)\text{I}_{1,1,1,1,1,1,0,0,0}\right.+\\
&+\left.  (p_4^2-t) \text{I}_{1,1,1,0,1,1,0,0,0}+t  \text{I}_{1,1,1,1,0,1,0,0,0}\right.+\\
&+\left. s (\text{I}_{0,1,1,1,1,1,0,0,0}- \text{I}_{1,1,0,1,1,1,0,0,0}- \text{I}_{1,1,1,1,1,1,0,-1,0}) \right)\,,\\
B_{69}&=\epsilon ^4 (p_4^2-t) \left(\text{I}_{1,1,1,1,1,1,0,-1,0}- \text{I}_{1,1,1,0,1,1,0,0,0}\right)\,,\\
B_{70}&= \epsilon ^4r_4 r_{13} \text{I}_{1,1,1,1,1,1,0,0,0} \,,\\
B_{71}&=\epsilon^3 r_1 r_5\left(2 \epsilon (p_4^2-s-t)\text{I}_{1,1,1,1,1,1,0,0,0} +(p_4^2-s-t) \text{I}_{1,1,1,1,0,2,0,0,0} \right.+\\
&+\left.  (p_4^2-t)\text{I}_{1,2,1,1,1,1,0,-1,0} -s  \text{I}_{1,2,1,0,1,1,0,0,0}\right)\,.
\end{align*}
In addition, we made the following choice of basis for the elliptic sectors,
\begin{align*}
B_{72}&=\epsilon ^4 s r_2  \text{I}_{0,1,1,1,1,1,1,0,0} \,,\\
B_{73}&= \epsilon ^4 s \text{I}_{0,1,1,1,1,1,1,0,-1} \,,\\
B_{74}&= \epsilon ^3 s^2 \text{I}_{0,2,1,1,1,1,1,0,0} \,,\\
B_{75}&=\epsilon ^4 s r_2  \text{I}_{0,1,1,1,1,1,2,0,0} \,,\\
B_{76}&=\epsilon ^4  s r_2 \text{I}_{1,1,0,1,1,1,1,0,0} \,,\\
B_{77}&= \epsilon ^4 s \text{I}_{1,1,-1,1,1,1,1,0,0} \,,\\
B_{78}&=\epsilon ^3 s^2  \text{I}_{2,1,0,1,1,1,1,0,0} \,,\\
B_{79}&=\epsilon ^4 s r_2  \text{I}_{1,1,0,1,2,1,1,0,0} \,,\\
B_{80}&=\epsilon ^4 s  \left(\text{I}_{1,1,1,1,1,1,1,-1,0} (s-p_4^2)+\text{I}_{1,1,1,1,1,1,1,-2,0}\right) \,,\\
B_{81}&=\frac{1}{2} \epsilon ^4 s  \left( (s-p_4^2)\text{I}_{1,1,1,1,1,1,1,0,-1} +t \text{I}_{1,1,1,1,1,1,1,-1,0}+2 \text{I}_{1,1,1,1,1,1,1,-1,-1}\right) \,,\\
B_{82}&=\epsilon ^4 r_2 r_4 r_{10}  \left((p_4^2-s)\text{I}_{1,1,1,1,1,1,1,0,0} -\text{I}_{1,1,1,1,1,1,1,-1,0}\right) \,,\\
B_{83}&=\epsilon ^4 r_2 r_6  \left((s-p_4^2)\text{I}_{1,1,1,1,1,1,1,0,-1} -t \text{I}_{1,1,1,1,1,1,1,-1,0}\right) \,,\\
B_{84}&=-\epsilon ^4 r_2 r_3 r_9  \text{I}_{1,1,1,1,1,1,1,-1,0} \,.
\end{align*}
The factors labelled by $\{r_i\}$ are the following square roots,
\begin{align}
\label{eq:famFroots}
\def\arraystretch{1.3}
\begin{array}{ll}
r_1= \sqrt{-p_4^2}\,, &r_2= \sqrt{-s}\,, \\[0.2 em]
r_3= \sqrt{-t}\,, & r_4= \sqrt{-p_4^2+s+t}\,, \\[0.2 em]
r_5= \sqrt{4 m^2-p_4^2}\,, &r_6= \sqrt{4 m^2-s}\,, \\[0.2 em]
r_7= \sqrt{4 m^2-t}\,, &r_8= \sqrt{4 m^2-p_4^2+s+t}\,, \\[0.2 em]
r_9= \sqrt{4 m^2 \left(p_4^2-s-t\right)+s t}\,, &r_{10}= \sqrt{4 m^2 s+t \left(p_4^2-s-t\right)}\,, \\[0.2 em]
r_{11}=\sqrt{4 m^2 t+p_4^2 s-s^2-s t}\,, & r_{12}= \sqrt{4 m^2 s \left(-p_4^2+s+t\right)-p_4^4 t}\,, \\[0.2 em]
r_{13}= \sqrt{-4 m^2 s t+p_4^4 (s+t)-p_4^6}\,,\quad \quad \quad & r_{14}=\sqrt{m^4 (-s)+2 m^2 t \left(-2 p_4^2+s+2 t\right)-s t^2}\,,\\[0.2 em]
\end{array}
\end{align}
\vspace{-38 pt}
\begin{align*}
\hspace{-110 pt} r_{15}= \sqrt{m^4 (-s)+2 m^2 (s+2 t) \left(-p_4^2+s+t\right)-s \left(-p_4^2+s+t\right){}^2}\,.
\end{align*}
The labelling has been chosen such that the radicands of the roots are irreducible polynomials. In the basis elements of the polylogarithmic sectors, namely $B_1,\ldots,B_{71}$, the 15 roots only appear in the following 11 combinations,
\begin{align}
 \left\{r_2 r_6,r_1 r_5,r_3 r_7,r_4 r_8,r_2 r_3 r_9,r_2 r_4 r_{11},r_3 r_4 r_{10},r_2 r_{14},r_2 r_{15},r_3 r_{12},r_4
   r_{13}\right\}\,.
\end{align}
It may also be verified that the same 11 combinations are sufficient to express all products of roots appearing in the letters. Hence, in principle is it possible to combine them and work with a reduced set of 11 independent square roots for the polylogarithmic sectors. 

In the choice of basis for the elliptic sectors, the root $r_2$  appears separately. Therefore, there are 12 independent combinations of roots in the full basis of the family.
\label{app:CanonicalBasis}

\section{Alphabet of the polylogarithmic sectors}
\label{app:AlphabetPolylogarithmic}
The full alphabet for the polylogarithmic sectors of family G is given by the following 76 letters,
\begin{align*}
\begin{array}{ll}
l_1= m^2\,,  & l_2= p_4^2\,, \\[0.2 em]
l_3= s\,, &l_4= t\,, \\[0.2 em]
l_5= s+t\,, &l_6= p_4^2-s\,, \\[0.2 em]
l_7= p_4^2-t\,, & l_8= -p_4^2+s+t\,, \\[0.2 em]
l_9= 4 m^2-p_4^2\,, & l_{10}= 4 m^2-s\,, \\[0.2 em]
l_{11}= 4 m^2-t\,, & l_{12}= 4 m^2-p_4^2+s+t\,, \\[0.2 em]
l_{13}= m^2 s+p_4^4-p_4^2 s\,, & l_{14}= m^2 p_4^2+s (s-p_4^2)\,, \\[0.2 em]
l_{15}= m^2 p_4^2+t (t-p_4^2)\,, & l_{16}= m^2(s+t)^2-p_4^2 s t\,, \\[0.2 em]
l_{17}= t (-p_4^2+s+t)-4 m^2 s\,, & l_{18}= -4 m^2 t-p_4^2 s+s^2+s t\,, \\[0.2 em]
l_{19}= 4 m^2(p_4^2-s-t)+s t\,, & l_{20}= p_4^4 t-4 m^2 s (-p_4^2+s+t)\,, \\[0.2 em]
l_{21}= m^2 p_4^2-(s+t) (p_4^2-s-t)\,, & l_{22}= -4m^2 s t-p_4^6+p_4^4 (s+t)\,, \\[0.2 em]
l_{23}= m^2 (p_4^2-t)^2+p_4^2 s (-p_4^2+s+t)\,, & l_{24}= m^2(p_4^2-s)^2+p_4^2 t (-p_4^2+s+t)\,, \\[0.3 em]
l_{25}= \dfrac{-p_4^2+r_1 r_5}{-p_4^2-r_1 r_5}\,, & l_{26}= \dfrac{-s+r_2 r_6}{-s-r_2 r_6}\,, \\[1 em]
l_{27}= \dfrac{-t+r_3 r_7}{-t-r_3 r_7}\,, &l_{28}= \dfrac{-p_4^2+2 s+r_1 r_5}{-p_4^2+2 s-r_1 r_5}\,, \\[1 em]
l_{29}= \dfrac{-p_4^2+2 t+r_1 r_5}{-p_4^2+2 t-r_1 r_5}\,, &l_{30}= \dfrac{-p_4^2+2 (s+t)+r_1 r_5}{-p_4^2+2 (s+t)-r_1 r_5}\,, \\[1 em]
l_{31}= \dfrac{s \left(p_4^2-2 m^2\right)+ p_4^2 r_2 r_6 }{s \left(p_4^2-2 m^2\right)-p_4^2 r_2 r_6}\,, &l_{32}= \dfrac{-t p_4^2+r_3 r_{12}}{t p_4^2-r_3 r_{12}}\,, \\[1 em]
l_{33}= \dfrac{2 p_4^2 m^2-2 t m^2-s p_4^2+s r_1 r_5}{2 p_4^2 m^2-2 t m^2-s p_4^2-s r_1 r_5}\,, &l_{34}= \dfrac{-4 m^2+p_4^2-s-t+r_4 r_8}{-4 m^2+p_4^2-s-t-r_4 r_8}\,, \\[1 em]
l_{35}= \dfrac{-2 (s+t) m^2+s p_4^2+s r_1 r_5}{-2 (s+t) m^2+s p_4^2-s r_1 r_5}\,, &l_{36}= \dfrac{2 \left(p_4^2-s\right) m^2-t p_4^2+t r_1 r_5}{2 \left(p_4^2-s\right) m^2-t p_4^2-t r_1 r_5}\,, \\[1 em]
l_{37}= \dfrac{-s t+r_2 r_3 r_9}{-s t-r_2 r_3 r_9}\,, &l_{38}= \dfrac{t \left(p_4^2-s-t\right)+r_3 r_4 r_{10}}{t \left(p_4^2-s-t\right)-r_3 r_4 r_{10}}\,, \\[1 em]
l_{39}= \dfrac{-4 \left(p_4^2-s-t\right) m^2-s t+r_2 r_7 r_9}{-4 \left(p_4^2-s-t\right) m^2-s t-r_2 r_7 r_9}\,, &l_{40}= \dfrac{s \left(p_4^2-s-t\right)+r_2 r_4 r_{11}}{s \left(p_4^2-s-t\right)-r_2 r_4 r_{11}}\,, \\[1 em]
l_{41}= \dfrac{-\left(4 m^2-s\right) p_4^2+r_1 r_2 r_5 r_6}{-\left(4 m^2-s\right) p_4^2-r_1 r_2 r_5 r_6}\,, &l_{42}= \dfrac{-\left(4 m^2-t\right) p_4^2+r_1 r_3 r_5 r_7}{-\left(4 m^2-t\right) p_4^2-r_1 r_3 r_5 r_7}\,, \\[1 em]
l_{43}= \dfrac{-s \left(m^2+t\right)+r_2 r_{14}}{-s \left(m^2+t\right)-r_2 r_{14}}\,, &l_{44}= \dfrac{-2 (s+t) m^2+s t+r_2 r_3 r_9}{-2 (s+t) m^2+s t-r_2 r_3 r_9}\,, \\[1 em]
l_{45}= \dfrac{-p_4^2 \left(4 m^2-p_4^2+s+t\right)+r_1 r_4 r_5 r_8}{-\left(4 m^2-p_4^2+s+t\right) p_4^2-r_1 r_4 r_5 r_8}\,, \quad \quad \quad \quad &l_{46}= \dfrac{s m^2+t \left(2 p_4^2-s\right)+r_2 r_{14}}{s m^2+t \left(2 p_4^2-s\right)-r_2 r_{14}}\,, \\[1 em]
\end{array}
\end{align*}
\begin{align*}
\begin{array}{ll}
l_{47}= \dfrac{-p_4^6+(s+t) p_4^4-2 m^2 s t+p_4^2 r_4 r_{13} }{-p_4^6+(s+t) p_4^4-2 m^2 s t-  p_4^2 r_4 r_{13}}\,,\quad \quad & \hspace{-30 pt} l_{48}= \dfrac{\left(-2 p_4^2+s+2 t\right) m^2-s t+r_2 r_{14}}{\left(-2 p_4^2+s+2 t\right) m^2-s t-r_2 r_{14}}\,, \\[1 em]
l_{49}= \dfrac{2 \left(-p_4^2+s+2 t\right) m^2-s t+r_3 r_6 r_9}{2 \left(-p_4^2+s+2 t\right) m^2-s t-r_3 r_6 r_9}\,, & \hspace{-30 pt}l_{50}= \dfrac{-\left(-2 p_4^2+s+4 t\right) m^2+s t+r_6 r_{14}}{-\left(-2 p_4^2+s+4 t\right) m^2+s t-r_6 r_{14}}\,, \\[1 em]
l_{51}= \dfrac{q_1+r_1 r_2 r_3 r_5 r_9}{q_1-r_1 r_2 r_3 r_5 r_9}\,, &\hspace{-30 pt}l_{52}= \dfrac{q_2+r_2 r_8 r_{11}}{q_2-r_2 r_8 r_{11}}\,, \\[1 em]
l_{53}= \dfrac{q_3+r_3 r_8 r_{10}}{q_3-r_3 r_8 r_{10}}\,, &\hspace{-30 pt}l_{54}= \dfrac{q_4+r_4 r_7 r_{10}}{q_4-r_4 r_7 r_{10}}\,, \\[1 em]
l_{55}= \dfrac{q_5+r_1 r_3 r_5 r_{12}}{q_5-r_1 r_3 r_5 r_{12}}\,, &\hspace{-30 pt}l_{56}= \dfrac{q_6+r_2 r_{15}}{q_6-r_2 r_{15}}\,, \\[1 em]
l_{57}= \dfrac{q_7+r_4 r_6 r_{11}}{q_7-r_4 r_6 r_{11}}\,, &\hspace{-30 pt}l_{58}= \dfrac{q_8+r_2 r_{15}}{q_8-r_2 r_{15}}\,, \\[1 em]
l_{59}= \dfrac{q_9+r_6 r_{15}}{q_9-r_6 r_{15}}\,, &\hspace{-30 pt}l_{60}= \dfrac{q_{10}+r_1 r_2 r_3 r_5 r_9}{q_{10}-r_1 r_2 r_3 r_5 r_9}\,, \\[1 em]
l_{61}= \dfrac{q_{11}+r_1 r_4 r_5 r_{13}}{q_{11}-r_1 r_4 r_5 r_{13}}\,, &\hspace{-30 pt}l_{62}= \dfrac{q_{12}+\left(s-p_4^2\right) r_3 r_4 r_{10}}{q_{12}-\left(s-p_4^2\right) r_3 r_4 r_{10}}\,, \\[1 em]
l_{63}= \dfrac{q_{13}+r_1 r_3 r_4 r_5 r_{10}}{q_{13}-r_1 r_3 r_4 r_5 r_{10}}\,, &\hspace{-30 pt}l_{64}= \dfrac{q_{14}+r_1  r_2 r_4 r_5 r_{11}}{q_{14}-r_1 r_2 r_4 r_5 r_{11}}\,, \\[1 em]
l_{65}= \dfrac{q_{15}+r_1 r_3 r_4 r_5 r_{10}}{q_{15}-r_1 r_3 r_4 r_5 r_{10}}\,, &\hspace{-30 pt}l_{66}= \dfrac{q_{16}+\left(p_4^2+t\right) r_2 r_4 r_{11}}{q_{16}-\left(p_4^2+t\right) r_2 r_4 r_{11}}\,, \\[1 em]
l_{67}= \dfrac{q_{17}+\left(m^2-p_4^2\right) r_2 r_{15}}{q_{17}-\left(m^2-p_4^2\right) r_2 r_{15}}\,, &\hspace{-30 pt}l_{68}= \dfrac{q_{18}+q_{19} r_1 r_2 r_4 r_5 r_{11}}{q_{18}-q_{19} r_1 r_2 r_4 r_5 r_{11}}\,, \\[1 em]
l_{69}=\left(\dfrac{q_{20}+2 r_2 r_9 r_{12}}{q_{20}-2 r_2 r_9 r_{12}} \right)\left( \dfrac{q_{21}+p_4^2 r_2 r_9 r_{12}}{q_{21}-p_4^2 r_2 r_9 r_{12}}\right)\,, \\[1 em]
l_{70}=\left(\dfrac{q_{22}+p_4^2 r_3 r_{10} r_{13}}{q_{22}-p_4^2 r_3 r_{10} r_{13}} \right) \left(\dfrac{q_{23}+2 r_3 r_{10} r_{13}}{q_{23}-2 r_3 r_{10} r_{13}}\right)\,, \\[1 em]
l_{71}=\left(\dfrac{q_{24}+p_4^2r_2 r_{11} r_{13}}{q_{24}-p_4^2 r_2 r_{11} r_{13}} \right) \left(\dfrac{q_{25}+2 r_2 r_{11} r_{13}}{q_{25}-2 r_2 r_{11} r_{13}}\right)\,, \\[1 em]
l_{72}= \left(\dfrac{q_{26}+2 r_4 r_{11} r_{15}}{q_{26}-2 r_4 r_{11} r_{15}} \right) \left(\dfrac{q_{27}+q_{28} r_4 r_{11} r_{15}}{q_{27}-q_{28} r_4 r_{11} r_{15}}\right)  \,, \\[1 em]
l_{73}=\left(\dfrac{q_{29}+p_4^2 r_4 r_{10} r_{12} }{q_{29}-p_4^2 r_4 r_{10} r_{12}} \right) \left(\dfrac{q_{30}+2 r_4 r_{10} r_{12}}{q_{30}-2 r_4 r_{10} r_{12}}\right)\,, \\[1 em]
l_{74}= \left(\dfrac{q_{31}+2 r_3 r_9 r_{14}}{q_{31}-2 r_3 r_9 r_{14}} \right) \left(\dfrac{q_{32}+\left(m^2+t\right) r_3 r_9 r_{14}}{q_{32}-\left(m^2+t\right) r_3 r_9 r_{14}}\right)\,, \\[1 em]
l_{75}=\left(\dfrac{q_{33}+2 r_2 r_3 r_4 r_9 r_{13}}{q_{33}-2 r_2 r_3 r_4 r_9 r_{13}} \right) \left(\dfrac{q_{34}+p_4^2 r_2 r_3 r_4 r_9 r_{13} }{q_{34}-p_4^2 r_2 r_3 r_4 r_9 r_{13}}\right)\,, \\[1 em]
l_{76}=\left(\dfrac{q_{35}+p_4^2 r_2 r_3 r_4 r_{11} r_{12} }{q_{35}-p_4^2 r_2 r_3 r_4 r_{11} r_{12}} \right) \left(\dfrac{q_{36}+2 r_2 r_3 r_4 r_{11} r_{12}}{q_{36}-2 r_2 r_3 r_4 r_{11} r_{12}}\right)\,, \\[1 em]
\end{array}
\end{align*}
where $q_i$ are the following polynomials:
\begin{align*}
q_1&= -s t p_4^2-4 m^2 \left(p_4^2-s-t\right) p_4^2\,, \\
q_2&= -2 (s-t) m^2-s \left(-p_4^2+s+t\right)\,, \\
q_3&= -2 (s-t) m^2-t \left(p_4^2-s-t\right)\,, \\
q_4&= -2 \left(p_4^2-2 s-t\right) m^2-t \left(-p_4^2+s+t\right)\,, \\
q_5&= 4 m^2 s \left(-p_4^2+s+t\right)-t p_4^4\,, \\
q_6&= s \left(-p_4^2+s+t\right)-m^2 (s+2 t)\,, \\
q_7&= -2 \left(p_4^2-s-2 t\right) m^2-s \left(-p_4^2+s+t\right)\,, \\
q_8&= -2 p_4^4+(3 s+2 t) p_4^2-s \left(m^2+s+t\right)\,, \\
q_9&= s \left(-p_4^2+s+t\right)-m^2 \left(-2 p_4^2+3 s+4 t\right)\,, \\
q_{10}&= -2 \left((s-t) p_4^2-s (s+t)\right) m^2-s t p_4^2\,, \\
q_{11}&= -p_4^6+(s+t) p_4^4-2 m^2 \left(-p_4^4+(s+t) p_4^2+s t\right)\,, \\
q_{12}&= -t \left(p_4^2+s\right) \left(p_4^2-s-t\right)\,, \\
q_{13}&= -2 \left((s-t) p_4^2+t (s+t)\right) m^2-t p_4^2 \left(p_4^2-s-t\right)\,, \\
q_{14}&= -2 \left((s-t) p_4^2-s (s+t)\right) m^2-s p_4^2 \left(-p_4^2+s+t\right)\,, \\
q_{15}&= -2 \left(-p_4^4+(s+t) p_4^2+s t\right) m^2-t p_4^2 \left(p_4^2-s-t\right)\,, \\
q_{16}&= -\left(p_4^2-t\right) \left(4 t m^2-s^2+s p_4^2-s t\right)\,, \\
q_{17}&= -s m^4-\left(s^2+t s+2 t p_4^2\right) m^2-s p_4^2 \left(p_4^2-s-t\right)\,, \\
q_{18}&= 2 \left((t-s) p_4^6+\left(s^2-5 t s-2 t^2\right) p_4^4+t \left(6 s^2+5 t s+t^2\right) p_4^2+s t^2 (s+t)\right) m^2 \\ 
      & \quad +s p_4^2 \left(p_4^6-(3 s+t) p_4^4+(2 s+t)^2 p_4^2-2 s^3-t^3-3 s t^2-4 s^2 t\right)\,, \\
q_{19}&= -\left(p_4^2-2 s-t\right) \left(p_4^2+t\right)\,, \\
q_{20}&= -8 s \left(-p_4^2+s+t\right) m^2-t \left(-p_4^4-s^2\right)\,, \\
q_{21}&= s t p_4^4+2 m^2 \left(p_4^2-s-t\right) \left(p_4^4+s^2\right)\,, \\
q_{22}&= t \left(p_4^2-s-t\right) p_4^4+2 m^2 s \left(p_4^4+t^2\right)\,, \\
q_{23}&= p_4^6-(s+t) p_4^4+t^2 p_4^2-t \left(t (s+t)-8 m^2 s\right)\,, \\
q_{24}&= -s p_4^6+\left(-2 t m^2+s^2+s t\right) p_4^4-2 m^2 s^2 t\,, \\
q_{25}&= -p_4^6+(s+t) p_4^4-s^2 p_4^2+s \left(-8 t m^2+s^2+s t\right)\,, \\
q_{26}&= -s m^4-2 (s+4 t) \left(p_4^2-s-t\right) m^2-2 s \left(-p_4^2+s+t\right){}^2\,, \\
q_{27}&= -2 t m^6+(s+4 t) \left(p_4^2-s-t\right) m^4+2 (s+2 t) \left(-p_4^2+s+t\right){}^2 m^2-s \left(-p_4^2+s+t\right){}^3\,, \\
q_{28}&= -m^2-p_4^2+s+t\,, \\
q_{29}&= t \left(p_4^2-s-t\right) p_4^4+2 m^2 s \left(2 p_4^4-2 (s+t) p_4^2+(s+t)^2\right)\,, \\
q_{30}&= t \left(2 p_4^4-2 (s+t) p_4^2+(s+t)^2\right)-8 m^2 s \left(-p_4^2+s+t\right)\,, \\
q_{31}&= -s m^4+2 t \left(-4 p_4^2+3 s+4 t\right) m^2-2 s t^2\,, \\
q_{32}&= 2   \left(-p_4^2+s+t\right) m^6+t \left(-4 p_4^2+3 s+4 t\right) m^4+2 t^2 \left(-2 p_4^2+s+2 t\right) m^2-s t^3\,, \\
q_{33}&= p_4^8-2 (s+t) p_4^6+(s+t)^2 p_4^4+8 m^2   s t p_4^2+s t \left(s t-8 m^2 (s+t)\right)\,, \\
q_{34}&= -s t \left(p_4^2-s-t\right) p_4^4-2 m^2 \left(p_4^8-2 (s+t) p_4^6+(s+t)^2 p_4^4+s^2 t^2\right)\,, \\
q_{35}&=   -s t \left(p_4^2-s-t\right) p_4^4-2 m^2 \left(\left(s^2+t^2\right) p_4^4-2 s^2 (s+t) p_4^2+s^2 (s+t)^2\right)\,, \\
q_{36}&= \left(s^2+t^2\right) p_4^4-2 s   \left(-4 t m^2+s^2+s t\right) p_4^2+s (s+t) \left(-8 t m^2+s^2+s t\right)\,.
\end{align*}

\bibliographystyle{JHEP}
\bibliography{refs}

\end{document}